\documentclass[floats,floatfix,showpacs,amssymb,prd,twocolumn,superscriptaddress,nofootinbib,nolongbibliography,reprint]{revtex4-2}

\usepackage{amssymb,amsmath,verbatim,mathtools,needspace,enumitem,etoolbox,graphicx,physics,microtype,afterpage,bigints,gensymb,tabularx,xspace,bbm}

\usepackage[dvipsnames, usenames]{xcolor}
\definecolor{linkcolor}{rgb}{0.0,0.3,0.5}
\usepackage[unicode, colorlinks=true, linkcolor=linkcolor, citecolor=linkcolor, filecolor=linkcolor,urlcolor=linkcolor, pdfusetitle]{hyperref}
\usepackage[all]{hypcap}
\usepackage[T1]{fontenc}
\usepackage[utf8]{inputenc}
\usepackage{orcidlink}
\usepackage{bibunits}
\usepackage{overpic,makecell,multirow} 
\usepackage{xtab,afterpage,longtable}
\usepackage[normalem]{ulem}

\newcommand{\ssim}{\mathchar"5218\relax\,}

\makeatletter
\newcommand*{\balancecolsandclearpage}{\close@column@grid \cleardoublepage \twocolumngrid}
\makeatother

\newcolumntype{L}[1]{>{\raggedright\let\newline\\\arraybackslash\hspace{0pt}}m{#1}}
\newcolumntype{C}[1]{>{\centering\let\newline\\\arraybackslash\hspace{0pt}}m{#1}}
\newcolumntype{R}[1]{>{\raggedleft\let\newline\\\arraybackslash\hspace{0pt}}m{#1}}

\newcommand{\sm}{Supplemental Material\xspace}

\newcommand\prlsec[1]{\vspace{2mm}\noindent \textbf{\textit{#1}}\,---}

\newcommand{\milan}{\affiliation{Dipartimento di Fisica ``G. Occhialini'', Universit\`a degli Studi di Milano-Bicocca, Piazza della Scienza 3, 20126 Milano, Italy}}
\newcommand{\infn}{\affiliation{INFN, Sezione di Milano-Bicocca, Piazza della Scienza 3, 20126 Milano, Italy}}

\makeatletter
\def\maketitle{
\@author@finish
\title@column\titleblock@produce
\suppressfloats[t]}
\makeatother

\begin{document}

\newcommand{\titleprl}{Which is which? Identification of the two compact objects in gravitational-wave binaries}
\title{\titleprl}

\author{Davide Gerosa$\,$\orcidlink{0000-0002-0933-3579}}
\email{davide.gerosa@unimib.it}

\milan \infn

\author{Viola De Renzis$\,$\orcidlink{0000-0001-7038-735X}}
\milan \infn

\author{Federica Tettoni$\,$\orcidlink{0009-0008-6460-2322}}
\milan

\author{Matthew Mould$\,$\orcidlink{0000-0001-5460-2910}}
\affiliation{LIGO, Massachusetts Institute of Technology, 77 Massachusetts Avenue, Cambridge, MA 02139, USA}
\affiliation{Kavli Institute for Astrophysics and Space Research \& Department of Physics,
Massachusetts Institute of Technology, 77 Massachusetts Avenue, Cambridge, MA 02139, USA}

\author{Alberto Vecchio$\,$\orcidlink{0000-0002-6254-1617}}
\affiliation{School of Physics and Astronomy \& Institute for Gravitational Wave Astronomy,
University of Birmingham, Birmingham, B15 2TT, United Kingdom}

\author{Costantino Pacilio$\,$\orcidlink{0000-0002-8140-4992}}
\milan \infn


\date{\today}

\begin{abstract}

Compact objects observed in gravitational-wave astronomy so far always come in pairs and never individually.
Identifying the two components of a binary system is a delicate operation that is often taken for granted. The labeling procedure (i.e., which is object ``1'' and which is object ``2'') effectively acts as systematics, or, equivalently an unspecified prior, in gravitational-wave data inference. The common approach is to label the objects solely by their masses, on a sample-by-sample basis. %
We show that object identification %
can instead be tackled using the posterior distribution as a whole. We frame the problem in terms of constrained clustering ---a flavor of semi-supervised machine learning--- and find that unfolding the labeling systematics can significantly impact, and arguably improve, our interpretation of the data. In particular, the precision of black-hole spin measurements improves by up to 50\%, 
multimodalities and tails tend to disappear, posteriors become closer to Gaussian distributions, and the identification of the nature of the object (i.e. black hole vs. neutron star) is facilitated. We estimate that about 10\% of the LIGO/Virgo posterior samples are affected by this relabeling, i.e. they might have been attributed to the other compact object in the observed~binaries.

\end{abstract}

\maketitle

\defaultbibliography{semisup}
\defaultbibliographystyle{apsrev4-2}
\begin{bibunit}

\prlsec{Labeling uncertainty}
The main sources of gravitational waves (GWs) are 
merging binaries made of black holes (BHs) and neutron stars (NS)~\cite{2019PhRvX...9c1040A,2021PhRvX..11b1053A,2024PhRvD.109b2001A,2023PhRvX..13d1039A}. 
Each GW signal thus contains joint information on two compact objects. %
While the GW signal itself is the same regardless of how we label the binary components, the (astro)physical exploitation of GW data often %
makes use of parameters describing single objects (masses, spins, tidal deformabilities, etc.).
Some examples include identifying if a GW event contains BHs or NSs (and thus whether it needs to be flagged for possible electromagnetic counterparts), measuring compact objects with unusual masses or spins (which is important for inferring their astrophysical formation pathway), and population-level inference (which is typically performed on the component properties). %

Untangling the properties of 
the two binary members 
in a GW signal 
is a non-trivial operation, even though it is often taken for granted. The standard approach is that of identifying the objects solely by their mass: one labels the heavier object in a binary by, e.g., ``1'' and the lighter object by, e.g., ``2.'' With this, one has that, for instance, $m_1$ is the mass of the heavier object, $m_2$ is the mass of the lighter object, $\chi_1$ is the spin of the heavier object, $\Lambda_2$ is the tidal deformability of the lighter object, etc. %
This becomes non-trivial when considering measurement errors. Say we observe a compact binary with true masses $m_1=10M_\odot$, $m_2=9M_\odot$ and a signal-to-noise ratio (SNR) such that the typical statistical error on the individual masses is $\Delta m \sim 2 M_\odot$. 
The error bars, or the posterior distributions, of the two masses overlap which implies one cannot confidently determine which object is heavier and which is lighter.
For cases where measurement errors ``swapped'' the two objects, our inference on, say, the spin $\chi_1$ will be polluted by information coming from the spin of the other object. In other words, the ``1'' and ``2'' labels %
are uncertain. This is a source of systematic, or equivalently an unspecified prior, %
in GW data analysis.

Some awareness on this point was first raised by \citeauthor{2021PhRvL.126q1103B} \cite{2021PhRvL.126q1103B}, %
who suggested using spin, rather than mass,
to label the components of a BH binary. With this assumption, one has $\chi_1>\chi_2$ and $m_1$ ($m_2$) becomes the mass of more (less) rapidly rotating BH. As correctly noted \cite{2021PhRvL.126q1103B}, this approach might fix the issue illustrated above for binaries with comparable masses but introduces spurious effects for binaries with well-measured masses but  overlapping spin posteriors (which is more likely, given the weaker effect the latter on the waveform). %

In this Letter, we tackle the labeling uncertainty in GW astronomy and show that %
data-driven labels improve the identifiability of the two binary components. Different GW events might be more appropriately described by different labelings, which implies the %
labeling procedure %
should depend on the data and not assumed a priori.

\prlsec{A machine-learning solution} 
Inference results in GW astronomy are provided in the form of Bayesian posterior samples. Suppose one has $N$ samples with $M$ variables for each of the binary components  (e.g. masses, spins, tidal deformabilities, etc.), totaling $N \times 2M$ real numbers. Each object can be labeled either ``1'' or ``2'' in each posterior sample, corresponding to $2^N$ possible labelings. The usual mass-sorting strategy corresponds to a specific choice among these many possibilities, namely, that where $m_1>m_2$ on a sample-by-sample basis. Reference~\cite{2021PhRvL.126q1103B} instead assumes $\chi_1>\chi_2$, again on a sample-by-sample basis. 
Among the available $2^N$ possibilities, can we choose a labeling that maximizes the identification of the two components of a GW binary?

This problem is akin to that of clustering in machine learning \cite{2015AnDS....2..165X,2024arXiv240107389Y,2024A&C....4800851F}, which tackles the identification of patterns in the data in an unsupervised fashion. Clustering algorithms aim at grouping the input datapoints according to their intrinsic similarities, which are defined through a suitable loss function. %
Let us now reshape the $N \times 2M$ matrix of posterior samples into a $2N \times M$ matrix by concatenating the features referring to the same parameters of the two components. %
For instance, considering the masses and spins of a BH binary (i.e. $M=2$), we would thus have $2N$ masses and $2N$ spins. A clustering strategy might then be employed to identify two clusters in the joint mass-spin space, which would then be associated with the two BHs of the binary. 

Clustering gets us half the way there but neglects a key piece of information. With a purely unsupervised approach, two objects from the same posterior sample might end up in the same cluster, which would be manifestly incorrect. Our clustering problem is not made of $2N$ independent datapoints, but rather of $N$ pairs where the two $M$-dimensional datapoints in each pair must belong to different clusters. In the machine-learning jargon, these are called ``cannot-link'' instances and are tackled with constrained clustering  \cite{Engelen:2020aa,Dinler2016,2023arXiv230300522G}  ---a form of semi-supervised learning where one has some knowledge of the targeted structure  (unlike unsupervised learning) but still cannot label individual datapoints (unlike supervised learning).

More concretely, spectral clustering \cite{2007arXiv0711.0189V} is a suitable algorithm for the problem at hand     %
because it is specifically designed to identify connectivity in the data, unlike most other clustering algorithms (e.g. K means, mixture models) which instead target compactness. In brief, one defines a $2N\times 2N$ positive and symmetric matrix $A$, hereafter the affinity matrix, that characterizes the connections between each pair of inputs. A typical choice is that of a radial-basis-function kernel
\begin{align}
A_{ij}  = \exp \left [ - \frac{d^2(x_i, x_j)}{2 \ell^2} \right]
\label{affinity}
\end{align}
where $x$ indicates our $M$-dimensional datapoints with $i,j=1,\dots 2N$, %
$d$ indicates a notion of distance in the feature space, and $\ell$ is a bandwidth parameter. The actual clustering happens in the space of the eigenvalues of a Laplacian projection of $A$; this allows tackling complex datasets where the targeted clusters cannot be suitably described by their first few moments. Larger (smaller) values of the affinity matrix indicate that those two datapoints are more (less) likely to belong to the same cluster. Cannot-link constraints can be implemented by zeroing out all %
 entries of the affinity matrix corresponding to objects coming from the same posterior samples. 

We use the \textsc{scikit-learn} \cite{2011JMLR...12.2825P} implementation of spectral clustering. As common in machine learning, we pre-process data by removing the mean and scaling to unit variance. One needs a kernel bandwidth that is orders of magnitude greater than the range of the pre-processed data for all constraints to be strictly enforced; in practice, we set $\ell=10^6$. As for the function $d$ entering Eq.~(\ref{affinity}), simple visual tests reveal that using the absolute sum (i.e. the $L_1$ norm) results in better performances than using the Euclidean distance. $L_1$ norms are commonly employed in machine learning to identify sparse solutions in both clustering~\cite{champion:hal-03095805,kawale2013constrained} and regression  \cite{Tibshirani:1996fxl}.
Our code is provided at Ref.~\cite{datarelease}; an injection test %
is presented in the \sm~\cite{suppmat}.

Once the two objects have been identified via clustering, one still has the freedom to associate the labels ``1'' and ``2'' to each of the two clusters. There is no ambiguity in this and all prescriptions are equivalent (because the objects have already been identified!). For concreteness, we refer to object 1 (object 2) as that with the larger (smaller) median value of the mass. Note the crucial difference here: while the usual strategy is that of labeling the two objects  \emph{a priori} %
on a sample-by-sample basis, we instead consider a summary statistics %
\emph{a posteriori}, i.e. after the full set of posterior samples has been looked at. 

In the following, we compare our semi-supervised approach against the usual sample-by-sample mass labeling by quoting the fraction $\zeta$ of posterior samples that are not relabeled. The parameter $\zeta$ thus ranges from 50\% (corresponding to randomly shuffled pairs) to 100\% (corresponding identical sets of labels).

\begin{figure*}

    \begin{overpic}[width=0.996\textwidth]{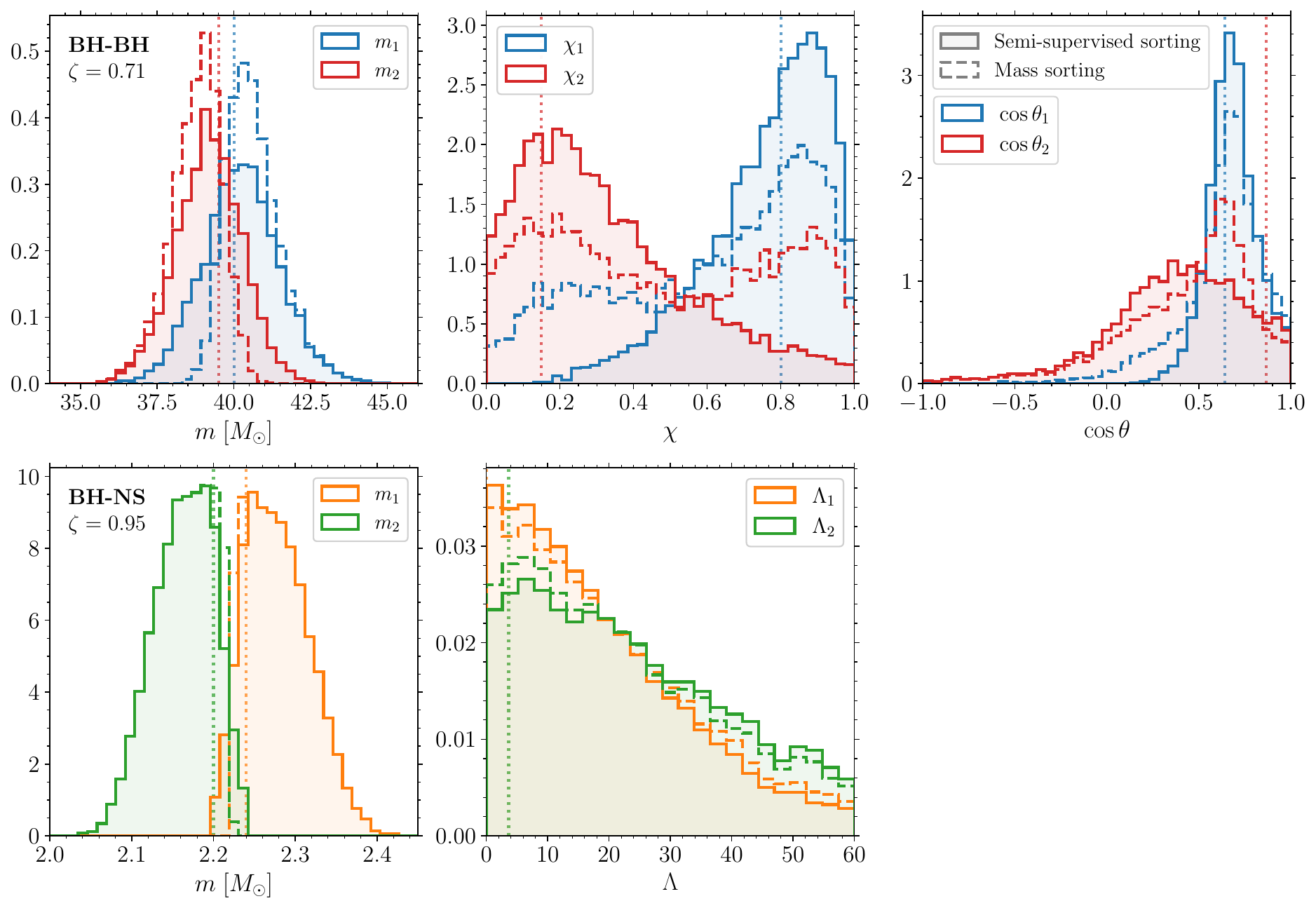}
        \put(67.1,19){
        \
\resizebox{5.4cm}{!}{
{   
\begin{tabular}{C{0.45cm}|C{1.3cm}|C{0.95cm}C{1.42cm}C{1.42cm}}
\multicolumn{2}{c|}{}& True& Mass & Semi-sup.\\[-0.03cm]
\multicolumn{2}{c|}{}& values& sorting & sorting\\[0.1cm]
\hline\hline
\rule{0pt}{0.35cm}
\multirow{3}{*}{$\!\!$\rotatebox{90}{\textbf{BH-BH\hspace{0.85cm}}}}
&$m_1\;[M_\odot]$&$40.0$&$40.5^{+1.8}_{-1.1}$&$40.2^{+2.1}_{-2.1}$\\ \rule{0pt}{0.4cm}
&$m_2\;[M_\odot]$&$39.5$&$38.9^{+1.1}_{-1.6}$&$39.2^{+1.7}_{-1.9}$\\ \rule{0pt}{0.4cm}
&$\chi_1$&$0.80$&$0.67^{+0.27}_{-0.56}$&$0.80^{+0.17}_{-0.36}$\\ \rule{0pt}{0.4cm}
&$\chi_2$&$0.15$&$0.45^{+0.49}_{-0.40}$&$0.28^{+0.51}_{-0.24}$\\ \rule{0pt}{0.4cm}
&$\cos\theta_1$&$0.64$&$0.67^{+0.26}_{-0.51}$&$0.68^{+0.24}_{-0.22}$\\ \rule{0pt}{0.4cm}
&$\cos\theta_2$&$0.87$&$0.53^{+0.36}_{-0.77}$&$0.41^{+0.50}_{-0.70}$
\\[0.15cm]
\hline
\rule{0pt}{0.35cm}
\multirow{2}{*}{$\!\!$\rotatebox{90}{\textbf{BH-NS\hspace{0.4cm}}}}
&$m_1\;[M_\odot]$&$2.24$&$2.28^{+0.07}_{-0.05}$&$2.28^{+0.07}_{-0.05}$\\ \rule{0pt}{0.4cm}
&$m_2\;[M_\odot]$&$2.20$&$2.17^{+0.05}_{-0.07}$&$2.17^{+0.05}_{-0.07}$\\ \rule{0pt}{0.4cm}
&$\Lambda_1$&$0.0$&$17.7^{+39.6}_{-16.2}$&$16.1^{+37.4}_{-14.7}$\\ \rule{0pt}{0.4cm}
&$\Lambda_2$&$3.6$&$21.0^{+44.8}_{-18.9}$&$22.9^{+45.0}_{-20.6}$\\
\end{tabular}
        }}
}
\renewcommand{\arraystretch}{1}
    \end{overpic}
\caption{Posterior distributions of two synthetic GW signals detectable by LIGO/Virgo. The top row shows masses $m$, spin magnitudes $\chi$, and spin directions $\cos\theta$ of a BH-BH system. The bottom row shows masses $m$ and tidal deformabilities $\Lambda$ of a mixed BH-NS binary. Colors indicate the two binary components identified using both the usual mass-sorting approach (dashed, empty histograms) and our clustering procedure (solid, filled histograms). The quantity $\zeta$ %
indicates the fraction of samples that are left invariant under the two labelings. The table to the right reports the injected values as well as medians and 90\% confidence intervals of the marginalized posterior distributions.}
\label{twoinj}
\end{figure*}

\begin{figure*}
\includegraphics[width=\textwidth]{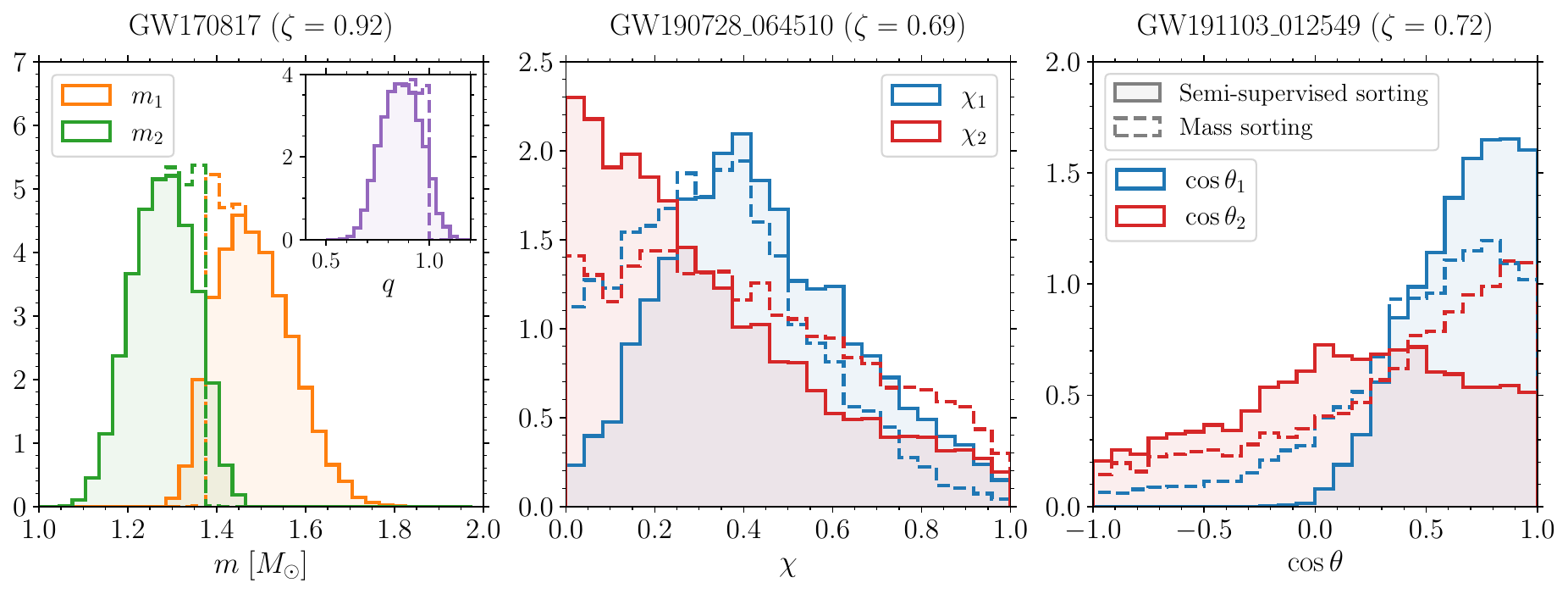}
\caption{Marginalized posterior distributions of representative events from current GW data. The left, middle, and right panel show the masses $m$ of binary NS GW170817, the spin magnitudes $\chi$ of binary BH GW190728$\_$064510, and the spin directions $\cos\theta$ of binary BH  GW191103$\_$012549, respectively.  The inset in the left panel shows the marginalized posterior distribution of the mass ratio $q=m_2/m_1$. Colors indicate the two binary components identified using both the usual mass-sorting approach (dashed, empty histograms) and our clustering procedure (solid, filled histograms). The quantity $\zeta$ %
indicates the fraction of samples that are left invariant under the two labelings. }
\label{someevents}
\end{figure*}

\begin{figure}
\includegraphics[width=\columnwidth]{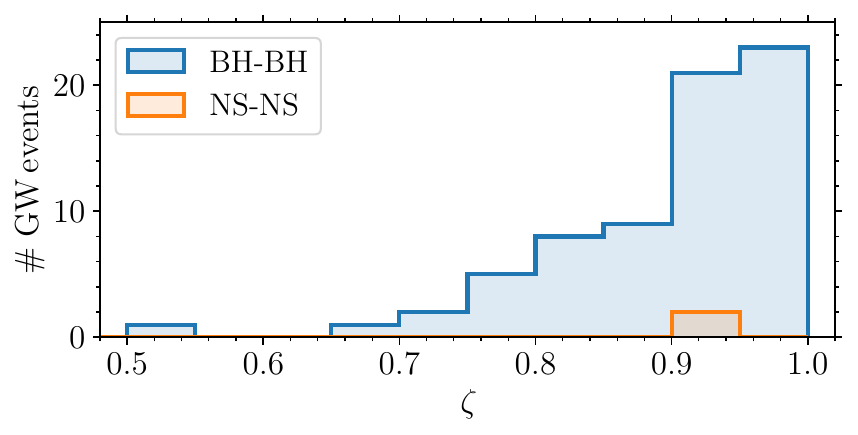}
\caption{Semi-supervised identification applied to BH binaries (blue) and NS binaries (orange) up to \mbox{GWTC-3}. Histograms show the fraction $\zeta$ of posterior samples that remain invariant in our proposed approach against the common mass-sorting strategy. Distributions are normalized to the number of events considered, which is 70 for binary BHs and 2 for binary NSs.}
\label{relabelandratios}
\end{figure}

\prlsec{Examples} Two GW examples are shown in Fig.~\ref{twoinj}. These are synthetic signals injected into the LIGO/Virgo network with SNR of $\ssim 45$; details are provided in the \sm~\cite{suppmat}.

The first example is a compact binary made of two BHs with similar masses ($m=40 M_\odot,39.5 M_\odot$ in the detector frame) but spins that are different in both magnitude ($\chi=0.8,0.15$) and direction ($\theta=50^\circ, 30^\circ$, where $\theta$ is the angle between the spin and the orbital angular momentum of the binary). Assuming the common sample-by-sample mass labeling, one finds that both spin magnitudes have bimodal posterior distributions. This is because inference on each spin is contaminated by the spin of the other BH. In turn, this translates into large error bars on both quantities: in line with the conventional wisdom, one might think that BH spins are poorly measured by LIGO/Virgo. Instead, this is partly due to the a-priori choice of labels.

 We employ our semi-supervised machine-learning strategy on $M=3$ features $(m,\chi,\cos\theta)$ and $N=10^4$ posterior samples. We use $\cos\theta$ instead of $\theta$ because the underlying prior is uniform in the former and have verified that considering the azimuthal angle $\phi$ does not impact the results presented below.  We use the detector-frame mass $m$ instead of the source-frame mass because the former is directly contained in the GW phase.  For the BH binary shown in Fig.~\ref{twoinj}, we find $\zeta\simeq 71\%$. %
 Crucially, the bimodality in the spin posteriors %
 disappears when objects are relabeled via clustering. Quoting the width of the 90\% credible interval, we find that the precision of the spin-magnitude measurements goes from $\Delta\chi\sim 0.85$ to $\sim 0.5$ for $\chi_1$ and $\sim 0.7$ for $\chi_2$, which is a notable improvement considering these parameters are defined in $[0,1]$. For the spin tilts $\theta$, the common mass-sorting approach results in posteriors of the angles $\theta_1$ and $\theta_2$ that both peak at the true value of $\theta_1$. This is due to the underlying labeling assumption and, indeed, the 
 peak in $\theta_2$ disappears when labels are sorted via clustering and the posterior moves closer to a uniform distribution. 

Our second example is a non-spinning system composed of a BH and a NS. The two objects have similar detector-frame masses $m =  2.24 M_\odot, 2.20 M_\odot$ but only the lighter object has a non-zero tidal deformability {$\Lambda =  3.6$} as predicted by a realistic equation of state~\cite{1998PhRvC..58.1804A}. Inferring whether any of the two components has a value of $\Lambda$ that is different from zero would imply that this GW signal contains a NS, with potential consequences for electromagnetic counterparts %
and the properties of matter at supranuclear densities~\cite{2024PhRvD.110f3014G}. %
We run the clustering algorithm on $M=2$ features $(m,\Lambda)$ and find that our inference on the nature of the object depends on the labeling strategy. %
Relabeling by clustering reveals that the tidal-deformability posterior of the NS (BH) has less (more) weight in the region close to $\Lambda= 0$ compared to the usual  mass-sorting assumption. We find $P(\Lambda_1<20) \sim 59\%$ (55\%) and $P(\Lambda_2<20) \sim 44\%$ (48\%) assuming clustering- (mass-) labels.
Note this example is rather conservative, $\zeta\simeq 95\%$. %

Such improvements in either spins or tidal deformabilities are \emph{not}  reflected in a loss of accuracy on the masses, with posteriors that remain well centered on the true values. The mass measurements obtained after relabeling are less precise though (i.e. posteriors are broader, their 90\% credible intervals increase by up to a factor of 1.5). The commonly quoted mass distributions are systematically more narrow because the mass itself was used to label the samples.

\prlsec{Current data} We consider 70 BH binaries and 2 NS binaries detected during the first three observing runs of LIGO/Virgo with false-alarm rate $< 1/{\rm yr}$ ~\cite{2019PhRvX...9c1040A,2021PhRvX..11b1053A,2024PhRvD.109b2001A,2023PhRvX..13d1039A}. We do not consider BH-NS events because they have been analyzed assuming $\Lambda_1=0$ and their masses are unequal.  As above, the clustering algorithm is exposed to $M=3$ features ($m,\chi,\cos\theta$) for binary BHs and $M=2$ features ($m,\Lambda$) for binary NSs, using at most $N=10^4$ posterior samples for computational efficiency. Some illustrative cases are presented in Fig.~\ref{someevents}; %
additional results are reported in the \sm~\cite{suppmat}. %

Figure~\ref{relabelandratios} shows the distributions of the relabeling fractions $\zeta$ across the GW catalog; this has a mean value of $\ssim 90\%$ and a standard deviation of $\ssim 8\%$. %
 The implication of our findings is that about 10\% of the posterior samples used in countless downstream applications in GW astronomy %
 might be better represented by a different labeling. About $35\%$ of the events have $\zeta<90\%$. As expected, events with mass ratio that confidently departs from unity such as GW190412, GW190814, and GW190929$\_$012149 have $\zeta\sim100\%$, i.e. the two sets of labels are equivalent.  %
Conversely, events with mass ratio close to unity are more severely impacted by our procedure; the most extreme case is GW200209$\_$085452 which has $\zeta \simeq 53\%$. %

Relabeling increases the precision of the BH spin measurements. For instance, 
Fig.~\ref{someevents} shows how the primary BH spin magnitude $\chi_1$ of GW190728$\_$064510 exhibits a pronounced peak at $\sim 0.4$ when samples are sorted via clustering. In particular, the value of the posterior at $\chi_1=0$ is about a factor of $\ssim 5$ smaller compared to the usual interpretation of the data. 
 Those posterior samples with low spin magnitude are absorbed by $\chi_2$, which is now distinctly peaked at $\ssim 0$.

Using %
GW191103$\_$012549 as an example, Fig.~\ref{someevents} illustrate a trend in the spin directions which is present in several of the events in the GW catalog, namely that posterior tails tend to disappear. When using the usual mass-sorting labels, the primary spin of GW191103$\_$012549 has a probability of being anti-aligned $P(\cos\theta_1<0)\sim 13\%$. This drops to $\ssim 0.1\%$ when one identifies the two BHs via clustering, i.e. we are 2 orders of magnitude more certain that the primary spin %
 is preferentially aligned to the binary orbital angular momentum, with potential consequences for the astrophysical formation channel of this event~\cite{2018PhRvD..98h4036G,2016ApJ...832L...2R,2015ApJ...810...58S}.

The features we just highlighted are rather generic across the entire catalog. Quoting the 90\% credible intervals of the marginalized posteriors, we find that the precision of the spin measurements increases by up to $\ssim 20\%$ in the magnitudes $\chi$ and $\ssim 50\%$ in the directions $\cos\theta$. On the other hand, some of the mass posterior distributions broaden by up to $\ssim 50\%$. But lower precision does not imply lower accuracy and the controlled experiments presented above and in the \sm~\cite{suppmat} suggest that the latter is not \mbox{impacted.} %

Interestingly, the mass posteriors obtained after relabeling are substantially more similar to Gaussian distributions for both binary BHs and binary NSs; this point is illustrated in Fig.~\ref{someevents} for GW170817  (see also Fig.~\ref{twoinj}). Gaussian posterior distributions are predicted in the large-SNR regime and this is a key assumption underpinning e.g. the use of Fisher matrices %
as a fast approximation to Bayesian pipelines~\cite{2008PhRvD..77d2001V}. We run the Anderson-Darling test of Gaussianity~\cite{anderson1952asymptotic} on all marginalized mass posteriors and find that the medians of the related $A^2$ statistic across the GW catalog decrease from $\ssim 77$ to $\ssim 48$ for $m_1$ and from $\ssim 43$ to $\ssim 32$ for $m_2$ when using clustering. %
The usual mass-sorting strategy %
truncates samples at the $m_1=m_2$ boundary, a feature which is most evident when considering marginalized distributions of the mass ratio $q=m_2/m_1$ (see Fig.~\ref{someevents}, but the same applies to the injections of Fig.~\ref{twoinj}). %
After relabeling, one has that $q$ is mostly but not strictly $<1$ and distributions extend smoothly into the $q>1$ region.

\prlsec{Outlook} GW sources consist of pairs of compact objects. Separating the two components requires a piece of information that is not directly present in the data and, as such, necessitates assumptions. %
The usual mass-sorting approach is intuitive because masses are the GW observables of each individual object that are typically best measured, though this intuition fails when the binary mass ratio is compatible with unity. Instead of operating on individual samples, we illustrate how the identification of %
compact objects in GW binaries can instead be tackled using the full posterior.  In particular, looking for compactness in the feature space may lead to an insightful representation of the properties of the individual objects, at the cost of losing the
ability to describe this identification a priori. Relabeling samples corresponds to identifying a suitable reparametrization, the functional form of which crucially depends on the data (and thus varies across the parameter space).
While the labeling problem for multiple signals has been studied extensively inside and outside of GW astronomy~\cite{10.1214/088342305000000016,2019PhRvD.100h4041B}, this Letter refers to identifying the two objects of a single source, which is conceptually different. 

Semi-supervised machine learning provides a practical and affordable solution to this problem. %
Running spectral clustering is a cheap operation, taking $\ssim$minutes on a laptop for $N=10^4$ posterior samples. At the same time, the time complexity of the algorithm is $O(N^3)$, which is rather unforgiving (but see e.g. Refs.~\cite{chen2010parallel,cai2014large,he2018fast,huang2019ultra} for optimization strategies).

Relabeling only affects the properties of individual objects. Quantities such as the chirp mass $M_{\rm chirp}$, the effective tidal deformability $\tilde \Lambda$, and the effective spin $\chi_{\rm eff}$ are manifestly invariant under the ``1'' $\longleftrightarrow$ ``2'' label exchange (unlike, e.g., the mass ratio $q$ and the precession parameter $\chi_{\rm p}$ as currently defined~\cite{2015PhRvD..91b4043S,2021PhRvD.103f4067G}). These ``binary estimators'' are by construction robust with respect to the labeling systematic highlighted in this Letter.

We are actively working toward developing a strategy to fold %
 our labeling procedure into GW population analyses. %
Methods to assess the information gain between prior and posterior also need to be investigated.

 Additional analyses are provided in the \sm~\cite{suppmat}, which includes Refs.~\cite{2015PhRvD..91d2003V,
2020MNRAS.493.3132S,
2019ApJS..241...27A,
2020MNRAS.498.4492S,
2021PhRvD.103j4056P,
2019PhRvD.100d4003D,
2020PhRvD.101l4059T,
2020LRR....23....3A,
2008PhRvD..78d4021R,
2018PhRvD..98h3007N}. 

\prlsec{Acknowledgements} 
We thank Alice Palladino for early exploratory work on this problem.
We thank
Ssohrab Borhanian,
Matteo Boschini,
Walter Del Pozzo,
Thomas Dent,
Cecilia Maria Fabbri, 
Giulia Fumagalli, 
Kostas Kritos,
Michele Mancarella, 
Christopher Moore,
Luca Reali, 
Arianna Renzini,
Stefano Rinaldi,  
Alexandre Toubiana,
and
Salvatore Vitale
for discussions.
We thank the LIGO/Virgo/KAGRA Collaboration for releasing public software and data products.
D.G., V.D.R., F.T., and C.P. are supported by
ERC Starting Grant No.~945155--GWmining, 
Cariplo Foundation Grant No.~2021-0555, 
MUR PRIN Grant No.~2022-Z9X4XS, 
MUR Grant ``Progetto Dipartimenti di Eccellenza 2023-2027'' (BiCoQ),
and the ICSC National Research Centre funded by NextGenerationEU. 
D.G. is supported by MSCA Fellowships
No.~101064542--StochRewind and No.~101149270--ProtoBH.
A.V. acknowledges the support of the Royal Society and the Wolfson Foundation.
Computational work was performed at CINECA with allocations 
through INFN and Bicocca.

\prlsec{Data availability} Supporting code is available at Ref.~\cite{datarelease}.

\putbib
\end{bibunit}

\begin{bibunit}
\setcounter{equation}{0}
\setcounter{figure}{0}
\setcounter{table}{0}
\clearpage \pagebreak
\renewcommand{\theequation}{S\arabic{equation}}
\renewcommand{\thefigure}{S\arabic{figure}}
\renewcommand{\bibnumfmt}[1]{[S#1]}
\renewcommand{\citenumfont}[1]{S#1}

\section*{\sm}

\begin{figure*}[]
\includegraphics[page=1,width=\textwidth]{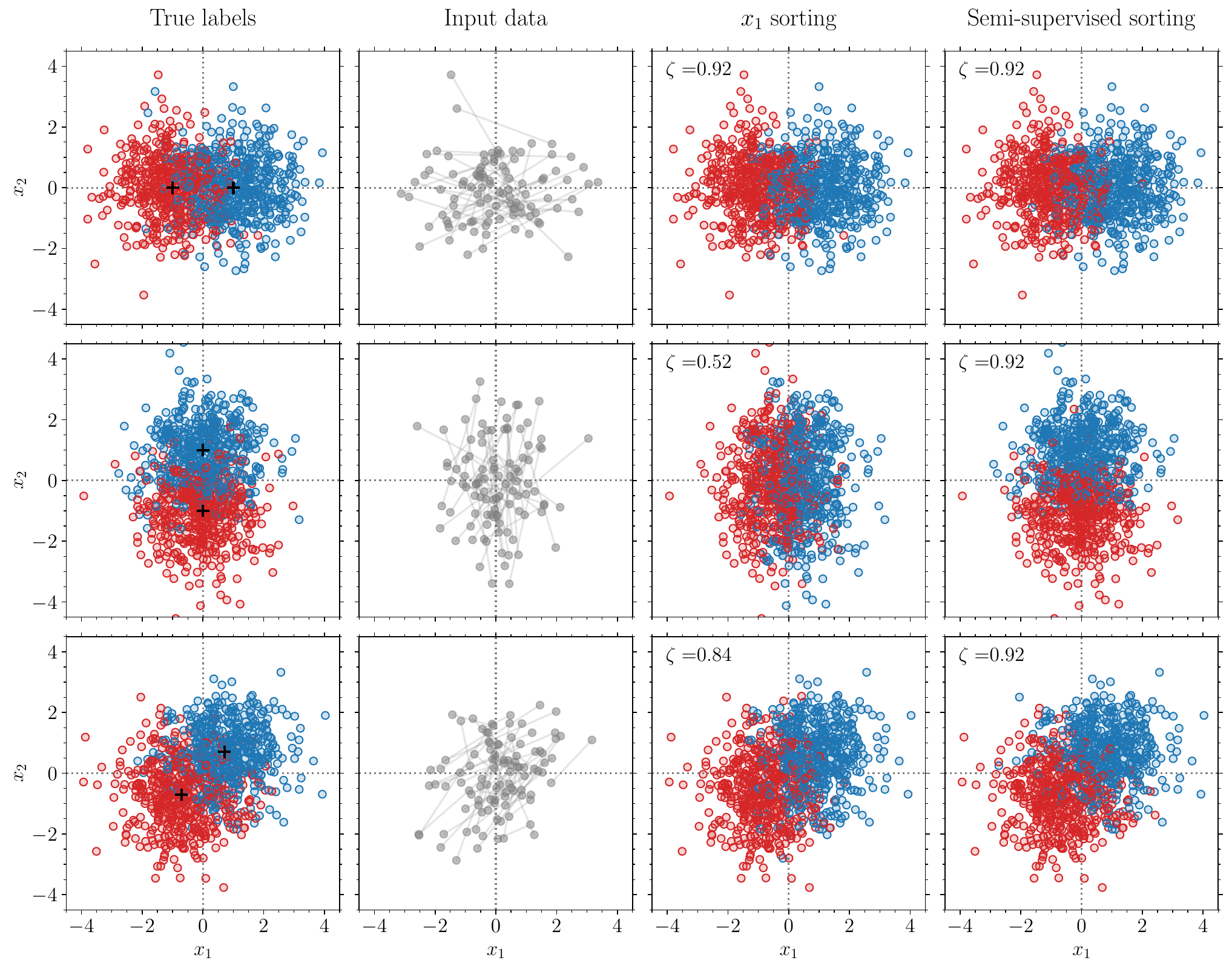}
\caption{Results from the test  %
 described in the \sm. The top, middle, and bottom rows assume $\delta\mu = (1,0), (1,0),$ and $(\sqrt{2}/2,\sqrt{2}/2)$, respectively. The two binary components are colored in blue and red. The underlying generating process is shown in the first column. We use uncorrelated Gaussian distributions with unit variance, their means are marked with black crosses. In reality, we do not know these labels but can only access a set of unlabeled pairs as shown in the second column (the distributions in those panels have been downsampled for readability). The usual approach is that of reconstructing the labels (i.e. the colors) using one of the features, for instance $x_1$, as shown in the third column. Results from our machine-learning strategy are illustrated in the fourth column. %
The quantity $\zeta$ indicates the fraction of true labels that have been correctly identified; the values of  $\zeta$ reported in this figure have been averaged over 100 realizations.}
\label{gausssteps}
\end{figure*}

\begin{figure*}
\includegraphics[page=1,width=\textwidth]{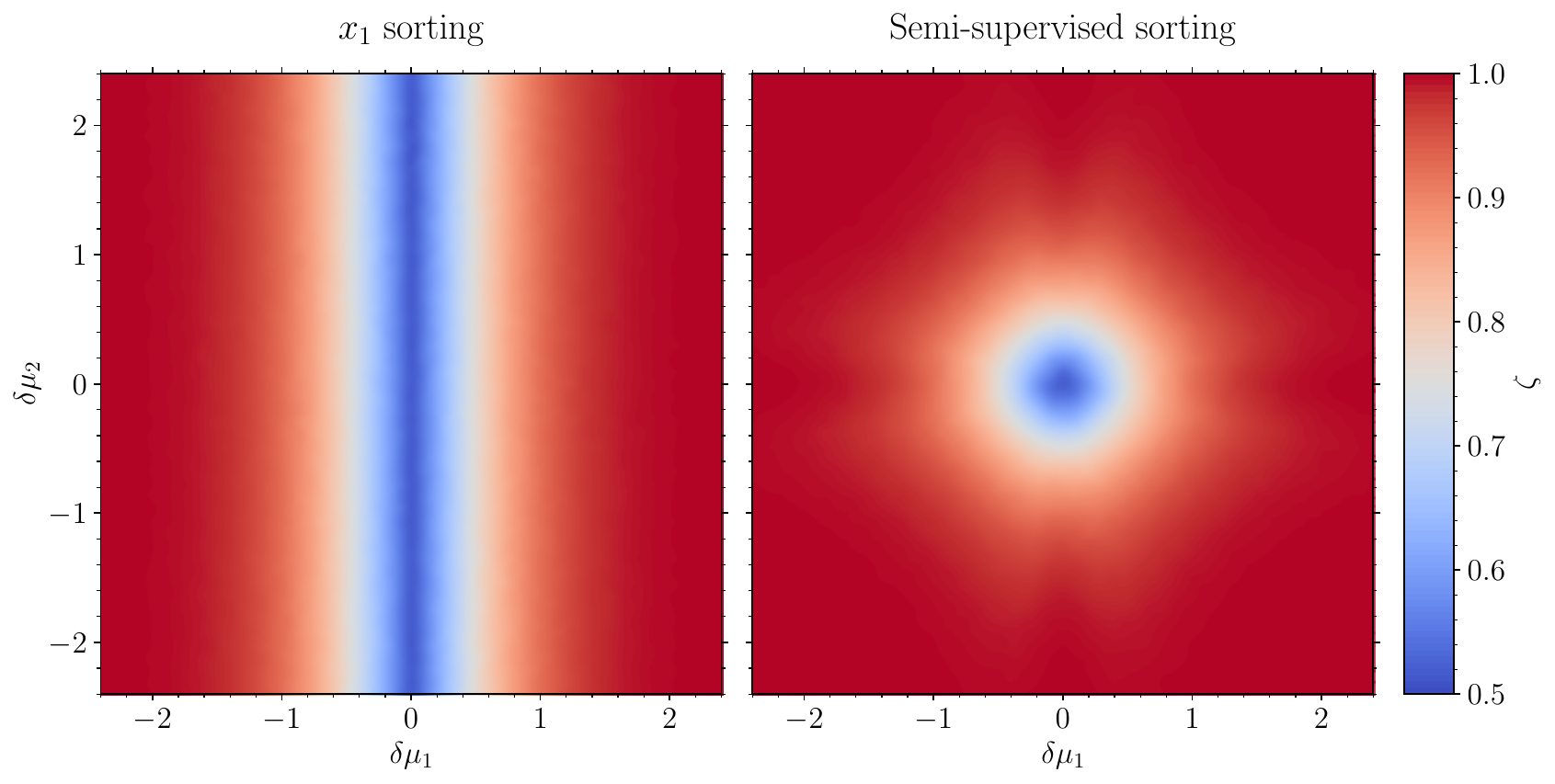}
\caption{Results from the test  %
 described in the \sm. We consider two bivariate, unit-variance Gaussian distributions separated by $\delta\mu=(\delta\mu_1,\delta\mu_2)$; these coordinates are reported on the $x$ and $y$ axes, respectively. The color scale indicates the fraction $\zeta$ of labels that are correctly identified, averaged over 100 realizations. In GW astronomy one usually sorts the labels using a single feature $x_1$ (left panel). The machine-learning strategy proposed in this paper (right panel) is consistently more accurate across the entire parameter space.}
\label{deltamuspace}
\end{figure*}

\prlsec{Test}
We present a simple test to highlight the key steps of our procedure. Much like injections are commonly used to test parameter-estimation pipelines before they are applied to real data, this example illustrates how a set of true labels is reconstructed with different labeling strategies. 

 In the following, the quantity $x_i^j$ is an $N$-dimensional vector where $N$ is the number of samples in the input posterior. The lower index $i=1,\dots, M$ labels the measured property while the upper index $j=1,2$ labels the object. %
Let us assume our data are normally distributed
with means $\mu^j$ and covariance matrices $\Sigma^j$. For simplicity, we assume that the two objects are measured equally well, i.e.   $\Sigma^1= \Sigma^2$.  Up to translations, rotations, and affine transformations, one can set  $\Sigma^1= \Sigma^2 = \mathbbm{1}
$, where $\mathbbm{1}$ is the identity matrix, and $\mu^1+\mu^2=0$. This toy model thus depends on a single free parameter $\delta\mu = (\mu^1 - \mu^2)/2$. Let us assume we have collected $N=500$ posterior samples describing $M=2$ properties. In the GW context, these could be the mass $(i=1)$ and spin $(i=2)$ of a BH binary.

Figure~\ref{gausssteps} shows three cases where the two %
distributions overlap at the $1\sigma$ level but along different directions. %
 This is a controlled experiment where we have access to the true labels $j$ that were used to generate the distributions (first column in Fig.~\ref{gausssteps}). The task at hand is to reconstruct those labels given only the set of unlabeled pairs (second column in Fig.~\ref{gausssteps}). The usual strategy employed in GW studies is that of labeling samples using one of the features, say $x_1$, on a sample-by-sample basis (third column in Fig.~\ref{gausssteps}). Instead, %
 we propose a semi-supervised machine-learning approach (fourth column~in~Fig.~\ref{gausssteps}).

 We present three configurations, quoting the fraction $\zeta \in[0.5,1]$ of the true labels that are correctly identified (note this is related but not equivalent to  the quantity $\zeta$ used in the main body of the paper, where we instead compare two alternative labeling strategies).
\begin{itemize}
\item The case with $\delta\mu = (1,0)$ is illustrated in top row of Fig.~\ref{gausssteps}. This might correspond to a GW event where the masses are distinct but the spins are not. In this case, sorting the labels using $x_1$ as commonly done is by construction the best strategy. We find that $\zeta=92\%$ of the true labels are correctly identified. Our semi-supervised approach identifies that the distinguishability power is entirely contained in $x_1$ are returns the same result.
\item In the middle row of Fig.~\ref{gausssteps}, we study the case with  $\delta\mu = (0,1)$. This models a BH binary with equal masses but different spins. Sorting the objects according to $x_1$ captures only $\zeta=52\%$ of the true labels while our clustering %
 strategy  still returns  $\zeta=92\%$. In this specific case, one might reach the same performance with a sample-by-sample sorting based on $x_2$, which is essentially the proposal of Ref.~\cite{2021PhRvL.126q1103B}.
\item Finally, the case with  $\delta\mu= (\sqrt{2}/2,\sqrt{2}/2)$ shown in the bottom row of Fig.~\ref{gausssteps} cannot be reduced to a simple sorting using either $x_1$ or $x_2$. Labeling the objects using $x_1$  identifies $\zeta=84\%$ of the pairs. The machine-learning approach is oblivious to these issues and still finds a solution with $\zeta=92\%$. 
\end{itemize}

Crucially and to facilitate comparisons, all three cases described above have the same norm  $|\delta\mu| = 1$.  The clustering algorithm recognizes that the problem is invariant under rotations  in the feature space %
and thus returns the same value of $\zeta$.  This point is further illustrated in Fig.~\ref{deltamuspace}, where we show the fraction of correctly identified labels $\zeta$ as a function of $\delta\mu$ for both the $x_1$ sorting  (left panel) and our semi-supervised sorting (right panel). %
Setting $\delta\mu=(0,0)$ in this toy model corresponds to two identical distributions; the underlying labels are fully degenerate (i.e. the two BHs are indistinguishable) and thus $\zeta\simeq 50\%$. With our spectral clustering algorithm, larger values of at least one component of $\delta\mu$ increase the labeling accuracy. The pattern shown in the right panel of Fig.~\ref{deltamuspace} reflects the radial symmetry of the problem. On the other hand, sorting by $x_1$ results in a more restrictive reflection symmetry about $\delta\mu_1=0$ which is a direct consequence of neglecting $x_2$ in the labeling process. 

In summary, our labeling strategy %
 can identify %
a sorting solution that is closer to the true labels, irrespective of the location of the source in the parameter space. %
The resulting values of $\zeta$ are consistently larger than those obtained when sorting according to one of the features. %
Repeating the exercise we just described with different input distributions (e.g. uniform distributions instead of Gaussians) returns the same trends.%

\prlsec{Gravitational-wave signals} %
The examples presented in Fig.~\ref{twoinj} make use of two synthetic GW signals from a BH-BH merger and a BH-NS merger on quasi-circular orbits. 
The BH-BH binary is injected with detector-frame masses $m_1=40 M_\odot$ and $m_2=39.5 M_\odot$, dimensionless spin magnitudes $\chi_{1}=0.80$ and $\chi_2=0.15$, tilt angles $\theta_{1}=50^\circ$ and $\theta_{2}=30^\circ$, and azimuthal angles $\phi_{12}=0.1$ and $\phi_{JL}=0.2$. The BH-NS system has $m_1=2.24 M_\odot$, $m_2=2.20 M_\odot$, and zero spins; the lighter object has tidal deformability $\Lambda_2=3.6$ as predicted by the ``APR4'' equation of state~\cite{1998PhRvC..58.1804A} while  $\Lambda_1=0$. The luminosity distance for both systems is set to obtain a network SNR of $\ssim 45$, resulting in $D_{\rm L}=551.5$ Mpc for the BH-BH  and {$D_{\rm L}=53$ Mpc} for the BH-NS.
All other parameters are the same for both injections; we set sky location $\alpha=0.5$ and $\delta=0.5$, direction of the line of sight $\theta_{JN}=80.2^\circ$, polarization $\psi=1$, time and phase of coalescence $t_{c}=0$ (in GPS time) and $\phi_{c}=\pi/4$~\cite{2015PhRvD..91d2003V}. 
For the BH-BH system, we perform statistical inference on all 15 parameters. For the BH-NS binary, we instead restrict our search to 9 parameters ($m_1$, $m_2$, $\Lambda_1$, $\Lambda_2$, $D_{\rm L}$, $\theta_{JN}$, $\psi$, $t_c$, $\phi_c$) for computational reasons related to the much longer signal. %
While this should be improved upon, it also mimics the case where the sky location can be pinpointed by the detection of an electromagnetic counterpart. We use  the {\sc Dynesty}~\cite{2020MNRAS.493.3132S} nested sampler wrapped by the {\sc Bilby}  package~\cite{2019ApJS..241...27A,2020MNRAS.498.4492S}. %
For both injection and recovery, we use the {\sc IMRPhenomXPHM}~\cite{2021PhRvD.103j4056P} and {\sc IMRPhenomPv2\_NRTidal}~\cite{2019PhRvD.100d4003D} waveform models for the BH-BH and BH-NS binary, respectively.  We checked that assuming the {\sc IMRPhenomNSBH} \cite{2020PhRvD.101l4059T} approximant instead of {\sc IMRPhenomPv2\_NRTidal} for the injected BH-NS signal returns similar results. %
We consider a three-detector network made of LIGO Livingston, LIGO Hanford, and Virgo using the O4 predicted sensitivities reported in Ref.~\cite{2020LRR....23....3A}. We set a lower frequency cutoff and a reference frequency of 20 Hz, assume a sampling frequency of 2048 Hz, and zero noise. We use uninformative priors as commonly employed in standard LIGO/Virgo analyses~\cite{2019PhRvX...9c1040A,2021PhRvX..11b1053A,2024PhRvD.109b2001A,2023PhRvX..13d1039A}. For the BH-NS injection, we restrict our priors to detector-frame chirp masses $M_{\rm chirp}\in[1,2]M_\odot$, mass ratios $q\in[0.125,1]$ assuming uniform priors in the component detector-frame masses {$m_1,m_2\in[2.1,2.3]M_\odot$.} %
We adopt uniform priors on  $\Lambda_1,\Lambda_2\in[0,500]$.

\begin{figure}
\includegraphics[width=0.905\columnwidth]{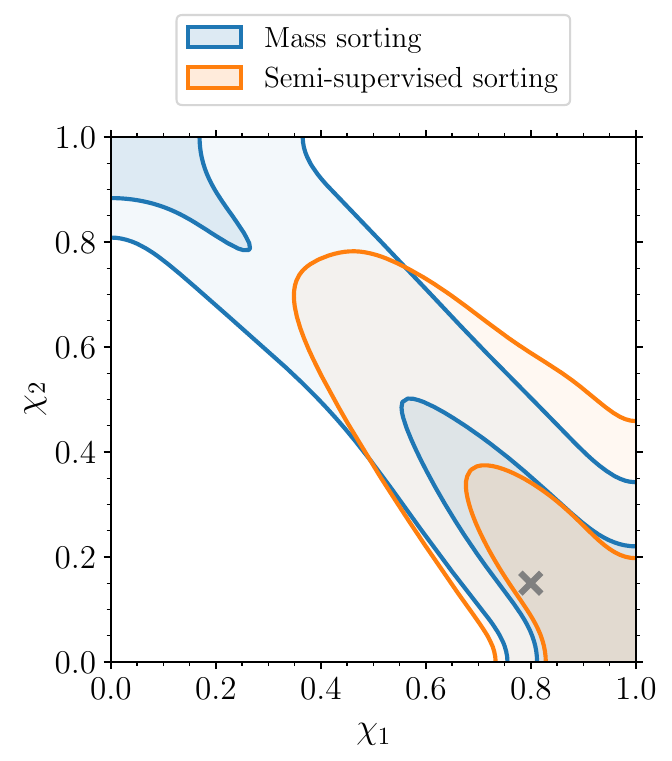}
\caption{Joint posterior distribution of the two spin magnitudes for our BH-BH injection (see Fig.~\ref{twoinj}). Blue (orange) contours refer to binary components identified using the usual mass-sorting approach (our clustering procedure); the grey cross marks the injected values. Contours refer to 50\% and 90\% credible intervals.}
\label{spinspin}
\end{figure}

For further insights, Fig.~\ref{spinspin} shows the joint  $\chi_1-\chi_2$ posterior distribution for our BH-BH injection. Marginalizing over either of the two spin magnitudes returns the histograms of Fig.~\ref{twoinj}. 
The strong anti-correlation between the two spins is a well-known feature in BH binary detections, which causes the effective spin $\chi_{\rm eff}$ to be better measured than any of the two individual spins~\cite{2008PhRvD..78d4021R,2018PhRvD..98h3007N}.  Qualitatively, the two-dimensional distribution of Fig.~\ref{spinspin} 
indicates that, in this source, ``one spin is large and the other one is small.''  The clustering labeling strategy ``folds'' the posterior, making this point evident also in the one-dimensional marginals of  Fig.~\ref{twoinj}. Crucially, this folding procedure is informed by the distribution itself and not decided a priori.

For Figs.~\ref{someevents} and \ref{relabelandratios}, 
we use posterior samples publicly released by the LIGO/Virgo/KAGRA Collaboration; for details see Refs.~\cite{2019PhRvX...9c1040A,2021PhRvX..11b1053A,2024PhRvD.109b2001A,2023PhRvX..13d1039A} and the related data releases.
 We consider 70 binary BHs and 2 binary NSs detected with FAR $<1/{\rm yr}$ in at least one of the pipelines,  see Table~\ref{table}. In particular:
\begin{itemize}
\item For binary BHs, we use samples released with GWTC-2.1 \cite{2024PhRvD.109b2001A} %
and \mbox{GWTC-3}~\cite{2023PhRvX..13d1039A}. We select the datasets labeled as ``cosmo'' and ``Mixed.'' 
\item For binary NS GW170817, we use the dataset labeled as ``IMRPhenomPv2NRT$\_$lowSpin'' from GWTC-1 \cite{2019PhRvX...9c1040A}.
\item For binary NS GW190425, we use the dataset labeled as ``cosmo'' and ``IMRPhenomPv2$\_$NRTidal:LowSpin'' from GWTC-2.1 \cite{2024PhRvD.109b2001A}.
 \end{itemize}
 
Note that we tackle the relabeling problem in post-processing; our conclusions are not affected by the parameter choice that was used during sampling.%

\begin{turnpage}
\capstartfalse
\begin{table*}
\vspace{-0.5cm}
\renewcommand{\arraystretch}{1.1}
\begin{tabular}{L{3cm}||C{0.8cm}|C{1.63cm}@{\hskip -0.1cm}C{1.63cm}|C{1.63cm}@{\hskip -0.1cm}C{1.63cm}|C{1.63cm}@{\hskip -0.1cm}C{1.63cm}|C{1.63cm}@{\hskip -0.1cm}C{1.63cm}|C{1.63cm}@{\hskip -0.1cm}C{1.63cm}|C{1.63cm}@{\hskip -0.1cm}C{1.63cm}}
&& Semi-sup. sorting & Mass sorting & Semi-sup. sorting & Mass sorting & Semi-sup. sorting & Mass sorting & Semi-sup. sorting & Mass sorting & Semi-sup. sorting & Mass sorting & Semi-sup. sorting & Mass sorting  \\[0.1cm]
\hline\hline
&&&&&&&&&&&\\[-0.3cm]
\multicolumn{1}{c||}{BH binary} & $\zeta$ & \multicolumn{2}{c|}{$m_1\;[M_\odot]$} & \multicolumn{2}{c|}{$m_2\;[M_\odot]$}  & \multicolumn{2}{c|}{$\chi_1$} & \multicolumn{2}{c|}{$\chi_2$}  & \multicolumn{2}{c|}{$\cos\theta_1$} & \multicolumn{2}{c}{$\cos\theta_2$} \\[0.1cm]
\hline
&&&&&&&&&&&\\[-0.3cm]
GW150914&$0.92$&$38.0^{+5.2}_{-3.3}$&$38.0^{+5.1}_{-2.9}$&$33.0^{+3.7}_{-5.1}$&$33.0^{+3.2}_{-5.0}$&$0.47^{+0.43}_{-0.42}$&$0.45^{+0.43}_{-0.41}$&$0.41^{+0.51}_{-0.36}$&$0.42^{+0.49}_{-0.37}$&$-0.24^{+0.82}_{-0.66}$&$-0.16^{+0.90}_{-0.72}$&$0.09^{+0.79}_{-0.91}$&$-0.04^{+0.88}_{-0.83}$\\ 
GW151012&$0.89$&$29.6^{+17.1}_{-9.9}$&$29.6^{+17.1}_{-6.9}$&$16.3^{+8.2}_{-6.0}$&$16.3^{+5.7}_{-6.0}$&$0.40^{+0.48}_{-0.36}$&$0.39^{+0.49}_{-0.34}$&$0.44^{+0.48}_{-0.39}$&$0.46^{+0.46}_{-0.41}$&$0.48^{+0.47}_{-0.97}$&$0.38^{+0.55}_{-1.08}$&$0.08^{+0.82}_{-0.93}$&$0.24^{+0.68}_{-1.06}$\\ 
GW151226&$0.85$&$15.4^{+12.5}_{-6.0}$&$15.6^{+12.4}_{-4.0}$&$8.2^{+5.1}_{-3.1}$&$8.1^{+2.6}_{-3.0}$&$0.66^{+0.29}_{-0.38}$&$0.61^{+0.34}_{-0.47}$&$0.37^{+0.52}_{-0.34}$&$0.45^{+0.47}_{-0.41}$&$0.58^{+0.36}_{-0.36}$&$0.52^{+0.39}_{-0.61}$&$0.04^{+0.85}_{-0.91}$&$0.18^{+0.74}_{-1.06}$\\ 
GW170104&$0.97$&$34.8^{+7.5}_{-5.1}$&$34.8^{+7.5}_{-4.9}$&$25.3^{+4.7}_{-5.5}$&$25.3^{+4.4}_{-5.5}$&$0.34^{+0.49}_{-0.31}$&$0.35^{+0.49}_{-0.32}$&$0.39^{+0.52}_{-0.35}$&$0.38^{+0.53}_{-0.34}$&$-0.23^{+0.95}_{-0.66}$&$-0.20^{+0.94}_{-0.69}$&$-0.00^{+0.88}_{-0.85}$&$-0.04^{+0.92}_{-0.82}$\\ 
GW170608&$0.81$&$11.3^{+4.4}_{-2.4}$&$11.4^{+4.3}_{-1.5}$&$8.4^{+2.3}_{-2.1}$&$8.4^{+1.2}_{-2.1}$&$0.30^{+0.44}_{-0.26}$&$0.29^{+0.45}_{-0.26}$&$0.31^{+0.52}_{-0.28}$&$0.31^{+0.52}_{-0.28}$&$0.38^{+0.54}_{-0.69}$&$0.22^{+0.67}_{-0.89}$&$-0.04^{+0.89}_{-0.83}$&$0.21^{+0.70}_{-1.02}$\\ 
GW170729&$0.92$&$77.8^{+18.2}_{-23.3}$&$77.9^{+18.1}_{-14.5}$&$44.9^{+22.8}_{-18.1}$&$44.0^{+18.1}_{-17.4}$&$0.64^{+0.31}_{-0.55}$&$0.60^{+0.35}_{-0.54}$&$0.43^{+0.49}_{-0.39}$&$0.48^{+0.46}_{-0.44}$&$0.71^{+0.27}_{-0.82}$&$0.67^{+0.31}_{-1.04}$&$0.16^{+0.75}_{-1.01}$&$0.26^{+0.69}_{-1.09}$\\ 
GW170809&$0.96$&$41.2^{+9.5}_{-6.7}$&$41.2^{+9.5}_{-6.3}$&$29.2^{+6.0}_{-6.5}$&$29.2^{+5.7}_{-6.5}$&$0.35^{+0.53}_{-0.32}$&$0.35^{+0.52}_{-0.32}$&$0.41^{+0.50}_{-0.37}$&$0.41^{+0.50}_{-0.37}$&$0.24^{+0.66}_{-0.99}$&$0.19^{+0.70}_{-0.98}$&$0.15^{+0.75}_{-0.99}$&$0.21^{+0.71}_{-1.03}$\\ 
GW170814&$0.86$&$34.6^{+6.2}_{-5.4}$&$34.7^{+6.1}_{-3.5}$&$28.2^{+4.7}_{-4.8}$&$28.1^{+3.2}_{-4.7}$&$0.44^{+0.46}_{-0.40}$&$0.41^{+0.48}_{-0.37}$&$0.39^{+0.52}_{-0.35}$&$0.42^{+0.50}_{-0.38}$&$0.40^{+0.52}_{-0.74}$&$0.28^{+0.62}_{-0.95}$&$-0.05^{+0.85}_{-0.82}$&$0.12^{+0.75}_{-0.95}$\\ 
GW170818&$0.94$&$42.2^{+8.0}_{-5.4}$&$42.2^{+8.0}_{-5.1}$&$33.5^{+5.4}_{-6.1}$&$33.4^{+5.0}_{-6.1}$&$0.52^{+0.42}_{-0.46}$&$0.52^{+0.42}_{-0.46}$&$0.47^{+0.47}_{-0.42}$&$0.47^{+0.47}_{-0.42}$&$-0.05^{+0.83}_{-0.75}$&$-0.11^{+0.86}_{-0.74}$&$-0.23^{+1.00}_{-0.69}$&$-0.16^{+0.96}_{-0.75}$\\ 
GW170823&$0.94$&$51.7^{+11.9}_{-8.4}$&$51.7^{+11.8}_{-7.9}$&$39.4^{+8.4}_{-11.0}$&$39.4^{+8.2}_{-11.0}$&$0.45^{+0.47}_{-0.40}$&$0.44^{+0.48}_{-0.39}$&$0.41^{+0.50}_{-0.37}$&$0.43^{+0.49}_{-0.39}$&$0.18^{+0.72}_{-0.94}$&$0.12^{+0.77}_{-0.93}$&$0.06^{+0.83}_{-0.91}$&$0.14^{+0.77}_{-0.96}$\\ 
GW190408$\_$181802&$0.96$&$31.7^{+7.0}_{-4.1}$&$31.7^{+7.0}_{-3.9}$&$23.8^{+3.8}_{-5.0}$&$23.8^{+3.6}_{-5.0}$&$0.29^{+0.52}_{-0.26}$&$0.31^{+0.51}_{-0.28}$&$0.39^{+0.50}_{-0.36}$&$0.37^{+0.52}_{-0.34}$&$-0.16^{+0.95}_{-0.73}$&$-0.14^{+0.93}_{-0.74}$&$-0.01^{+0.87}_{-0.85}$&$-0.03^{+0.89}_{-0.84}$\\ 
GW190412&$0.99$&$31.8^{+6.8}_{-6.6}$&$31.8^{+6.8}_{-6.6}$&$10.3^{+2.2}_{-1.6}$&$10.3^{+2.2}_{-1.6}$&$0.32^{+0.21}_{-0.19}$&$0.31^{+0.22}_{-0.20}$&$0.40^{+0.50}_{-0.37}$&$0.41^{+0.50}_{-0.37}$&$0.80^{+0.18}_{-0.47}$&$0.80^{+0.18}_{-0.54}$&$0.31^{+0.62}_{-1.08}$&$0.33^{+0.61}_{-1.09}$\\ 
GW190413$\_$052954&$0.90$&$52.8^{+15.2}_{-11.9}$&$52.8^{+15.2}_{-10.2}$&$38.2^{+10.6}_{-11.9}$&$37.9^{+9.9}_{-11.6}$&$0.38^{+0.52}_{-0.34}$&$0.42^{+0.50}_{-0.37}$&$0.49^{+0.44}_{-0.44}$&$0.45^{+0.48}_{-0.41}$&$0.03^{+0.83}_{-0.88}$&$-0.08^{+0.91}_{-0.80}$&$-0.22^{+1.04}_{-0.70}$&$-0.11^{+0.97}_{-0.79}$\\ 
GW190413$\_$134308&$0.94$&$83.0^{+19.6}_{-17.3}$&$83.0^{+19.6}_{-15.6}$&$50.5^{+20.6}_{-24.1}$&$50.5^{+18.8}_{-24.1}$&$0.64^{+0.32}_{-0.56}$&$0.60^{+0.36}_{-0.54}$&$0.47^{+0.47}_{-0.42}$&$0.50^{+0.44}_{-0.45}$&$-0.09^{+0.87}_{-0.77}$&$-0.06^{+0.87}_{-0.79}$&$0.09^{+0.81}_{-0.97}$&$0.04^{+0.86}_{-0.93}$\\ 
GW190421$\_$213856&$0.95$&$60.8^{+13.1}_{-8.9}$&$60.8^{+13.1}_{-8.7}$&$47.1^{+9.0}_{-14.5}$&$47.1^{+8.9}_{-14.5}$&$0.42^{+0.50}_{-0.38}$&$0.44^{+0.48}_{-0.40}$&$0.49^{+0.45}_{-0.44}$&$0.46^{+0.47}_{-0.42}$&$-0.20^{+0.98}_{-0.70}$&$-0.24^{+0.98}_{-0.67}$&$-0.28^{+1.06}_{-0.65}$&$-0.24^{+1.05}_{-0.69}$\\ 
GW190503$\_$185404&$0.94$&$53.1^{+12.1}_{-10.8}$&$53.2^{+12.0}_{-10.1}$&$36.5^{+11.1}_{-12.7}$&$36.5^{+10.1}_{-12.7}$&$0.43^{+0.47}_{-0.39}$&$0.43^{+0.47}_{-0.39}$&$0.45^{+0.49}_{-0.40}$&$0.44^{+0.49}_{-0.40}$&$-0.28^{+1.01}_{-0.63}$&$-0.21^{+0.99}_{-0.69}$&$0.08^{+0.81}_{-0.93}$&$-0.01^{+0.89}_{-0.87}$\\ 
GW190512$\_$180714&$0.97$&$29.3^{+6.8}_{-7.0}$&$29.3^{+6.8}_{-6.6}$&$15.8^{+5.0}_{-2.9}$&$15.8^{+4.6}_{-2.9}$&$0.19^{+0.46}_{-0.17}$&$0.20^{+0.49}_{-0.18}$&$0.43^{+0.49}_{-0.38}$&$0.41^{+0.50}_{-0.37}$&$0.10^{+0.79}_{-0.93}$&$0.08^{+0.79}_{-0.91}$&$0.10^{+0.79}_{-0.91}$&$0.12^{+0.78}_{-0.93}$\\ 
GW190513$\_$205428&$0.91$&$50.1^{+14.8}_{-16.9}$&$50.1^{+14.8}_{-12.8}$&$25.6^{+14.0}_{-7.0}$&$25.6^{+11.0}_{-7.0}$&$0.44^{+0.45}_{-0.40}$&$0.42^{+0.46}_{-0.37}$&$0.43^{+0.50}_{-0.39}$&$0.45^{+0.47}_{-0.41}$&$0.56^{+0.39}_{-0.94}$&$0.49^{+0.46}_{-1.09}$&$0.16^{+0.76}_{-1.00}$&$0.28^{+0.65}_{-1.10}$\\ 
GW190517$\_$055101&$0.94$&$52.9^{+15.0}_{-11.7}$&$52.9^{+15.0}_{-10.0}$&$32.8^{+11.2}_{-12.3}$&$32.8^{+9.5}_{-12.3}$&$0.90^{+0.08}_{-0.25}$&$0.90^{+0.09}_{-0.30}$&$0.59^{+0.36}_{-0.51}$&$0.62^{+0.35}_{-0.54}$&$0.80^{+0.18}_{-0.44}$&$0.79^{+0.19}_{-0.46}$&$0.40^{+0.55}_{-1.19}$&$0.41^{+0.54}_{-1.21}$\\ 
GW190519$\_$153544&$0.87$&$94.7^{+15.9}_{-33.2}$&$94.8^{+15.8}_{-12.4}$&$61.3^{+26.5}_{-20.0}$&$59.9^{+16.6}_{-19.0}$&$0.67^{+0.29}_{-0.42}$&$0.61^{+0.33}_{-0.46}$&$0.49^{+0.43}_{-0.45}$&$0.59^{+0.36}_{-0.52}$&$0.72^{+0.25}_{-0.57}$&$0.65^{+0.32}_{-0.70}$&$0.41^{+0.53}_{-1.08}$&$0.52^{+0.43}_{-1.15}$\\ 
GW190521&$0.93$&$151.3^{+32.6}_{-23.1}$&$152.2^{+31.7}_{-17.5}$&$90.0^{+59.4}_{-52.3}$&$89.8^{+48.9}_{-52.1}$&$0.73^{+0.24}_{-0.63}$&$0.71^{+0.27}_{-0.62}$&$0.50^{+0.44}_{-0.45}$&$0.53^{+0.42}_{-0.48}$&$-0.54^{+1.31}_{-0.41}$&$-0.50^{+1.34}_{-0.44}$&$0.21^{+0.71}_{-1.04}$&$0.14^{+0.76}_{-1.00}$\\
GW190521$\_$074359&$0.89$&$51.9^{+7.8}_{-8.0}$&$52.2^{+7.6}_{-5.4}$&$40.5^{+9.5}_{-7.3}$&$40.4^{+5.9}_{-7.2}$&$0.28^{+0.49}_{-0.26}$&$0.33^{+0.48}_{-0.30}$&$0.47^{+0.43}_{-0.42}$&$0.43^{+0.47}_{-0.38}$&$-0.00^{+0.82}_{-0.86}$&$0.13^{+0.74}_{-0.95}$&$0.50^{+0.45}_{-0.97}$&$0.42^{+0.53}_{-1.10}$\\ 
GW190527$\_$092055&$0.75$&$49.4^{+30.0}_{-21.4}$&$51.1^{+30.3}_{-9.3}$&$35.0^{+21.8}_{-15.4}$&$32.3^{+12.2}_{-13.2}$&$0.48^{+0.45}_{-0.41}$&$0.38^{+0.53}_{-0.34}$&$0.29^{+0.56}_{-0.27}$&$0.39^{+0.52}_{-0.36}$&$0.51^{+0.44}_{-0.76}$&$0.30^{+0.61}_{-1.02}$&$-0.07^{+0.88}_{-0.80}$&$0.23^{+0.69}_{-1.04}$\\ 
GW190602$\_$175927&$0.97$&$106.7^{+25.2}_{-18.8}$&$106.7^{+25.1}_{-18.1}$&$67.8^{+23.7}_{-32.6}$&$67.8^{+23.1}_{-32.6}$&$0.48^{+0.46}_{-0.43}$&$0.47^{+0.46}_{-0.42}$&$0.50^{+0.44}_{-0.45}$&$0.51^{+0.43}_{-0.46}$&$0.30^{+0.61}_{-0.98}$&$0.26^{+0.64}_{-0.97}$&$0.19^{+0.73}_{-1.03}$&$0.24^{+0.69}_{-1.06}$\\ 
GW190620$\_$030421&$0.83$&$86.7^{+24.3}_{-32.2}$&$87.1^{+23.9}_{-17.3}$&$55.2^{+24.9}_{-25.9}$&$53.3^{+17.6}_{-24.3}$&$0.77^{+0.21}_{-0.43}$&$0.70^{+0.27}_{-0.55}$&$0.46^{+0.47}_{-0.41}$&$0.57^{+0.39}_{-0.51}$&$0.72^{+0.25}_{-0.61}$&$0.64^{+0.33}_{-0.81}$&$0.35^{+0.59}_{-1.12}$&$0.50^{+0.46}_{-1.24}$\\ 
GW190630$\_$185205&$0.93$&$41.4^{+8.0}_{-7.8}$&$41.4^{+8.0}_{-6.5}$&$28.2^{+7.3}_{-5.6}$&$28.2^{+5.7}_{-5.6}$&$0.25^{+0.42}_{-0.22}$&$0.27^{+0.44}_{-0.25}$&$0.46^{+0.45}_{-0.41}$&$0.43^{+0.48}_{-0.38}$&$0.23^{+0.68}_{-1.00}$&$0.28^{+0.64}_{-1.01}$&$0.41^{+0.53}_{-1.04}$&$0.36^{+0.57}_{-1.07}$\\ 
GW190701$\_$203306&$0.97$&$74.5^{+16.1}_{-11.1}$&$74.6^{+16.1}_{-10.9}$&$56.1^{+12.4}_{-18.0}$&$56.1^{+12.1}_{-18.0}$&$0.44^{+0.48}_{-0.40}$&$0.44^{+0.49}_{-0.39}$&$0.45^{+0.49}_{-0.40}$&$0.45^{+0.48}_{-0.41}$&$-0.28^{+1.00}_{-0.64}$&$-0.24^{+0.99}_{-0.68}$&$-0.12^{+0.97}_{-0.78}$&$-0.17^{+1.00}_{-0.74}$\\ 
GW190706$\_$222641&$0.93$&$117.5^{+22.6}_{-33.0}$&$117.5^{+22.6}_{-18.6}$&$64.4^{+37.2}_{-28.1}$&$63.7^{+27.2}_{-27.4}$&$0.67^{+0.28}_{-0.50}$&$0.64^{+0.32}_{-0.51}$&$0.46^{+0.47}_{-0.42}$&$0.51^{+0.43}_{-0.46}$&$0.64^{+0.33}_{-0.72}$&$0.60^{+0.36}_{-0.88}$&$0.27^{+0.66}_{-1.08}$&$0.35^{+0.60}_{-1.15}$\\ 
GW190707$\_$093326&$0.97$&$14.1^{+3.0}_{-2.5}$&$14.1^{+3.0}_{-2.3}$&$9.3^{+1.8}_{-1.5}$&$9.3^{+1.7}_{-1.5}$&$0.19^{+0.44}_{-0.17}$&$0.19^{+0.46}_{-0.18}$&$0.35^{+0.52}_{-0.32}$&$0.34^{+0.53}_{-0.31}$&$-0.15^{+0.95}_{-0.74}$&$-0.17^{+0.95}_{-0.72}$&$-0.13^{+0.97}_{-0.73}$&$-0.10^{+0.95}_{-0.76}$\\ 
GW190708$\_$232457&$0.96$&$23.4^{+5.0}_{-5.4}$&$23.4^{+5.0}_{-5.1}$&$13.7^{+3.9}_{-2.2}$&$13.7^{+3.6}_{-2.2}$&$0.21^{+0.44}_{-0.19}$&$0.22^{+0.47}_{-0.20}$&$0.35^{+0.52}_{-0.32}$&$0.33^{+0.53}_{-0.30}$&$0.24^{+0.67}_{-0.93}$&$0.22^{+0.68}_{-0.91}$&$0.16^{+0.74}_{-0.97}$&$0.20^{+0.71}_{-1.01}$\\ 
GW190719$\_$215514&$0.77$&$57.0^{+76.9}_{-28.5}$&$59.1^{+74.8}_{-16.5}$&$34.9^{+25.1}_{-18.4}$&$32.5^{+14.5}_{-16.1}$&$0.69^{+0.27}_{-0.50}$&$0.60^{+0.35}_{-0.51}$&$0.41^{+0.51}_{-0.37}$&$0.52^{+0.42}_{-0.47}$&$0.66^{+0.31}_{-0.71}$&$0.57^{+0.40}_{-1.00}$&$0.13^{+0.78}_{-0.97}$&$0.33^{+0.62}_{-1.14}$\\ 
GW190720$\_$000836&$0.93$&$16.5^{+6.4}_{-5.2}$&$16.5^{+6.4}_{-3.9}$&$8.8^{+3.7}_{-2.1}$&$8.8^{+2.4}_{-2.1}$&$0.37^{+0.36}_{-0.31}$&$0.36^{+0.36}_{-0.31}$&$0.45^{+0.47}_{-0.41}$&$0.47^{+0.45}_{-0.42}$&$0.59^{+0.37}_{-0.83}$&$0.54^{+0.42}_{-1.03}$&$0.44^{+0.52}_{-1.17}$&$0.52^{+0.44}_{-1.20}$\\ 
GW190725$\_$174728&$0.82$&$14.0^{+12.7}_{-5.8}$&$14.2^{+12.4}_{-3.6}$&$7.7^{+5.3}_{-3.1}$&$7.6^{+2.4}_{-3.0}$&$0.33^{+0.51}_{-0.30}$&$0.35^{+0.52}_{-0.31}$&$0.52^{+0.43}_{-0.46}$&$0.49^{+0.44}_{-0.45}$&$0.19^{+0.72}_{-0.79}$&$-0.00^{+0.89}_{-0.82}$&$-0.36^{+1.09}_{-0.56}$&$-0.14^{+0.98}_{-0.75}$\\ 
GW190727$\_$060333&$0.84$&$58.5^{+13.5}_{-11.5}$&$58.9^{+13.1}_{-8.0}$&$46.8^{+10.7}_{-14.0}$&$46.2^{+8.3}_{-13.5}$&$0.57^{+0.37}_{-0.49}$&$0.50^{+0.43}_{-0.45}$&$0.39^{+0.52}_{-0.35}$&$0.46^{+0.47}_{-0.42}$&$0.39^{+0.54}_{-0.93}$&$0.24^{+0.67}_{-0.99}$&$-0.03^{+0.89}_{-0.84}$&$0.17^{+0.73}_{-1.01}$\\ 
GW190728$\_$064510&$0.69$&$13.7^{+9.0}_{-4.7}$&$14.5^{+8.1}_{-2.6}$&$10.0^{+5.4}_{-3.4}$&$9.4^{+1.9}_{-2.9}$&$0.41^{+0.41}_{-0.28}$&$0.33^{+0.38}_{-0.29}$&$0.25^{+0.57}_{-0.23}$&$0.37^{+0.51}_{-0.34}$&$0.68^{+0.29}_{-0.47}$&$0.49^{+0.46}_{-1.07}$&$0.08^{+0.78}_{-0.89}$&$0.43^{+0.52}_{-1.17}$\\ 
GW190731$\_$140936&$0.95$&$64.9^{+15.7}_{-11.5}$&$64.9^{+15.6}_{-10.8}$&$46.4^{+12.9}_{-17.7}$&$46.3^{+12.7}_{-17.5}$&$0.38^{+0.53}_{-0.35}$&$0.39^{+0.52}_{-0.35}$&$0.46^{+0.46}_{-0.42}$&$0.46^{+0.47}_{-0.42}$&$0.27^{+0.65}_{-1.06}$&$0.21^{+0.69}_{-1.03}$&$0.08^{+0.81}_{-0.94}$&$0.15^{+0.75}_{-1.00}$\\ 
GW190803$\_$022701&$0.97$&$57.6^{+13.1}_{-8.9}$&$57.6^{+13.1}_{-8.7}$&$43.0^{+9.3}_{-13.9}$&$43.0^{+9.3}_{-13.9}$&$0.40^{+0.51}_{-0.36}$&$0.41^{+0.50}_{-0.37}$&$0.46^{+0.47}_{-0.42}$&$0.45^{+0.48}_{-0.40}$&$-0.03^{+0.88}_{-0.82}$&$-0.06^{+0.89}_{-0.80}$&$-0.07^{+0.93}_{-0.82}$&$-0.04^{+0.90}_{-0.85}$\\ 
GW190805$\_$211137&$0.85$&$86.9^{+25.2}_{-24.8}$&$87.2^{+24.9}_{-15.3}$&$61.1^{+20.6}_{-24.6}$&$59.6^{+16.4}_{-23.3}$&$0.80^{+0.18}_{-0.50}$&$0.75^{+0.22}_{-0.59}$&$0.51^{+0.43}_{-0.45}$&$0.59^{+0.37}_{-0.52}$&$0.75^{+0.23}_{-0.69}$&$0.69^{+0.28}_{-0.94}$&$0.31^{+0.62}_{-1.09}$&$0.43^{+0.53}_{-1.18}$\\ 
GW190814&$1.00$&$24.5^{+1.6}_{-1.4}$&$24.5^{+1.6}_{-1.4}$&$2.7^{+0.1}_{-0.1}$&$2.7^{+0.1}_{-0.1}$&$0.03^{+0.07}_{-0.03}$&$0.03^{+0.07}_{-0.03}$&$0.48^{+0.45}_{-0.43}$&$0.48^{+0.45}_{-0.43}$&$0.00^{+0.94}_{-0.93}$&$0.00^{+0.94}_{-0.93}$&$0.11^{+0.81}_{-0.95}$&$0.11^{+0.81}_{-0.95}$

\end{tabular}

\end{table*}
\end{turnpage}

\begin{turnpage}
\begin{table*}[p]
\begin{tabular}{L{3cm}||C{0.8cm}|C{1.63cm}@{\hskip -0.1cm}C{1.63cm}|C{1.63cm}@{\hskip -0.1cm}C{1.63cm}|C{1.63cm}@{\hskip -0.1cm}C{1.63cm}|C{1.63cm}@{\hskip -0.1cm}C{1.63cm}|C{1.63cm}@{\hskip -0.1cm}C{1.63cm}|C{1.63cm}@{\hskip -0.1cm}C{1.63cm}}
&& Semi-sup. sorting & Mass sorting & Semi-sup. sorting & Mass sorting & Semi-sup. sorting & Mass sorting & Semi-sup. sorting & Mass sorting & Semi-sup. sorting & Mass sorting & Semi-sup. sorting & Mass sorting  \\[0.1cm]
\hline\hline
&&&&&&&&&&&\\[-0.3cm]
\multicolumn{1}{c||}{BH binary (cont.)} & $\zeta$ & \multicolumn{2}{c|}{$m_1\;[M_\odot]$} & \multicolumn{2}{c|}{$m_2\;[M_\odot]$}  & \multicolumn{2}{c|}{$\chi_1$} & \multicolumn{2}{c|}{$\chi_2$}  & \multicolumn{2}{c|}{$\cos\theta_1$} & \multicolumn{2}{c}{$\cos\theta_2$} \\[0.1cm]
\hline
&&&&&&&&&&&\\[-0.3cm]
GW190828$\_$063405&$0.79$&$42.7^{+7.9}_{-7.9}$&$43.3^{+7.4}_{-4.6}$&$36.2^{+6.7}_{-7.6}$&$35.5^{+4.7}_{-7.0}$&$0.53^{+0.40}_{-0.43}$&$0.45^{+0.47}_{-0.40}$&$0.32^{+0.55}_{-0.29}$&$0.41^{+0.50}_{-0.37}$&$0.60^{+0.36}_{-0.58}$&$0.46^{+0.48}_{-0.98}$&$0.03^{+0.82}_{-0.86}$&$0.28^{+0.64}_{-1.06}$\\ 
GW190828$\_$065509&$0.97$&$30.4^{+8.2}_{-8.7}$&$30.4^{+8.2}_{-8.3}$&$13.4^{+5.3}_{-2.7}$&$13.4^{+4.8}_{-2.7}$&$0.26^{+0.43}_{-0.23}$&$0.27^{+0.46}_{-0.24}$&$0.45^{+0.48}_{-0.41}$&$0.44^{+0.49}_{-0.40}$&$0.21^{+0.69}_{-0.94}$&$0.20^{+0.71}_{-0.93}$&$0.16^{+0.76}_{-1.00}$&$0.18^{+0.73}_{-1.03}$\\ 
GW190910$\_$112807&$0.91$&$56.5^{+8.8}_{-8.0}$&$56.5^{+8.7}_{-6.5}$&$44.6^{+8.5}_{-9.4}$&$44.5^{+7.5}_{-9.2}$&$0.30^{+0.53}_{-0.28}$&$0.32^{+0.52}_{-0.29}$&$0.38^{+0.52}_{-0.34}$&$0.36^{+0.54}_{-0.32}$&$0.15^{+0.74}_{-0.97}$&$0.04^{+0.82}_{-0.90}$&$-0.18^{+0.97}_{-0.73}$&$-0.07^{+0.93}_{-0.82}$\\ 
GW190915$\_$235702&$0.92$&$42.9^{+11.1}_{-6.8}$&$43.0^{+11.1}_{-5.9}$&$32.6^{+6.4}_{-8.1}$&$32.6^{+5.6}_{-8.0}$&$0.60^{+0.35}_{-0.52}$&$0.55^{+0.39}_{-0.49}$&$0.40^{+0.51}_{-0.37}$&$0.45^{+0.48}_{-0.41}$&$-0.02^{+0.76}_{-0.73}$&$-0.03^{+0.79}_{-0.76}$&$-0.12^{+0.98}_{-0.78}$&$-0.10^{+0.95}_{-0.79}$\\ 
GW190924$\_$021846&$0.93$&$9.8^{+4.8}_{-2.8}$&$9.8^{+4.8}_{-2.0}$&$5.7^{+2.1}_{-1.6}$&$5.7^{+1.3}_{-1.6}$&$0.22^{+0.45}_{-0.20}$&$0.23^{+0.46}_{-0.21}$&$0.35^{+0.52}_{-0.31}$&$0.33^{+0.53}_{-0.30}$&$0.19^{+0.70}_{-0.86}$&$0.10^{+0.77}_{-0.84}$&$0.06^{+0.84}_{-0.88}$&$0.17^{+0.74}_{-0.96}$\\ 
GW190925$\_$232845&$0.86$&$24.7^{+7.6}_{-4.7}$&$24.7^{+7.6}_{-3.2}$&$18.5^{+4.2}_{-4.1}$&$18.5^{+2.7}_{-4.0}$&$0.38^{+0.48}_{-0.34}$&$0.36^{+0.49}_{-0.32}$&$0.38^{+0.52}_{-0.35}$&$0.41^{+0.50}_{-0.37}$&$0.38^{+0.54}_{-0.87}$&$0.25^{+0.65}_{-0.97}$&$0.07^{+0.82}_{-0.90}$&$0.25^{+0.66}_{-1.01}$\\ 
GW190929$\_$012149&$0.99$&$101.8^{+26.1}_{-18.8}$&$101.8^{+26.1}_{-18.6}$&$41.8^{+25.0}_{-17.9}$&$41.8^{+24.6}_{-17.9}$&$0.33^{+0.54}_{-0.30}$&$0.34^{+0.54}_{-0.30}$&$0.48^{+0.45}_{-0.44}$&$0.48^{+0.46}_{-0.43}$&$-0.13^{+0.88}_{-0.76}$&$-0.12^{+0.88}_{-0.76}$&$-0.10^{+0.97}_{-0.81}$&$-0.10^{+0.97}_{-0.81}$\\ 
GW190930$\_$133541&$0.79$&$16.2^{+9.4}_{-7.8}$&$16.3^{+9.3}_{-4.5}$&$8.1^{+7.5}_{-2.4}$&$8.0^{+2.8}_{-2.4}$&$0.47^{+0.39}_{-0.36}$&$0.39^{+0.39}_{-0.35}$&$0.34^{+0.55}_{-0.31}$&$0.45^{+0.47}_{-0.40}$&$0.70^{+0.27}_{-0.53}$&$0.57^{+0.40}_{-1.02}$&$0.13^{+0.77}_{-0.94}$&$0.41^{+0.54}_{-1.16}$\\ 
GW191103$\_$012549&$0.72$&$13.4^{+8.0}_{-4.2}$&$14.0^{+7.4}_{-2.3}$&$9.8^{+4.6}_{-3.3}$&$9.4^{+1.8}_{-2.9}$&$0.56^{+0.36}_{-0.33}$&$0.46^{+0.40}_{-0.40}$&$0.34^{+0.55}_{-0.31}$&$0.50^{+0.44}_{-0.44}$&$0.69^{+0.27}_{-0.44}$&$0.54^{+0.41}_{-0.94}$&$0.18^{+0.73}_{-0.96}$&$0.48^{+0.48}_{-1.19}$\\ 
GW191105$\_$143521&$0.95$&$13.0^{+4.5}_{-2.0}$&$13.0^{+4.5}_{-1.8}$&$9.4^{+1.6}_{-2.2}$&$9.4^{+1.4}_{-2.2}$&$0.21^{+0.50}_{-0.20}$&$0.23^{+0.53}_{-0.21}$&$0.36^{+0.52}_{-0.33}$&$0.34^{+0.54}_{-0.31}$&$-0.03^{+0.87}_{-0.82}$&$-0.07^{+0.90}_{-0.78}$&$-0.06^{+0.92}_{-0.79}$&$-0.02^{+0.89}_{-0.83}$\\ 
GW191109$\_$010717&$0.91$&$80.5^{+13.1}_{-12.1}$&$81.1^{+12.7}_{-8.9}$&$60.3^{+20.0}_{-18.0}$&$59.9^{+15.6}_{-17.6}$&$0.85^{+0.13}_{-0.53}$&$0.83^{+0.15}_{-0.58}$&$0.60^{+0.36}_{-0.54}$&$0.65^{+0.31}_{-0.59}$&$-0.63^{+0.71}_{-0.34}$&$-0.60^{+0.87}_{-0.37}$&$-0.19^{+0.99}_{-0.72}$&$-0.27^{+1.04}_{-0.66}$\\ 
GW191127$\_$050227&$0.83$&$84.8^{+61.7}_{-46.1}$&$86.4^{+60.1}_{-37.3}$&$40.8^{+39.6}_{-27.2}$&$38.5^{+31.1}_{-25.3}$&$0.73^{+0.24}_{-0.58}$&$0.66^{+0.31}_{-0.58}$&$0.44^{+0.49}_{-0.40}$&$0.54^{+0.41}_{-0.49}$&$0.48^{+0.47}_{-0.83}$&$0.37^{+0.56}_{-1.00}$&$0.05^{+0.86}_{-0.91}$&$0.22^{+0.71}_{-1.06}$\\ 
GW191129$\_$134029&$0.95$&$12.3^{+4.9}_{-2.5}$&$12.3^{+4.9}_{-2.3}$&$7.8^{+1.9}_{-1.9}$&$7.8^{+1.7}_{-1.9}$&$0.23^{+0.35}_{-0.21}$&$0.25^{+0.37}_{-0.22}$&$0.37^{+0.50}_{-0.33}$&$0.34^{+0.52}_{-0.31}$&$0.28^{+0.63}_{-0.99}$&$0.27^{+0.64}_{-0.95}$&$0.29^{+0.64}_{-1.03}$&$0.31^{+0.63}_{-1.07}$\\ 
GW191204$\_$171526&$0.92$&$13.4^{+3.8}_{-2.7}$&$13.4^{+3.8}_{-2.0}$&$9.3^{+2.3}_{-1.8}$&$9.3^{+1.5}_{-1.8}$&$0.42^{+0.36}_{-0.36}$&$0.40^{+0.37}_{-0.35}$&$0.44^{+0.43}_{-0.39}$&$0.46^{+0.41}_{-0.40}$&$0.50^{+0.44}_{-0.69}$&$0.43^{+0.49}_{-0.89}$&$0.35^{+0.59}_{-1.06}$&$0.44^{+0.50}_{-1.07}$\\ 
GW191215$\_$223052&$0.97$&$33.5^{+9.3}_{-4.8}$&$33.5^{+9.3}_{-4.7}$&$24.5^{+4.2}_{-5.2}$&$24.5^{+4.1}_{-5.2}$&$0.48^{+0.43}_{-0.44}$&$0.47^{+0.44}_{-0.43}$&$0.43^{+0.50}_{-0.39}$&$0.44^{+0.49}_{-0.40}$&$-0.07^{+0.85}_{-0.67}$&$-0.09^{+0.87}_{-0.69}$&$-0.14^{+0.96}_{-0.76}$&$-0.11^{+0.94}_{-0.78}$\\ 
GW191216$\_$213338&$0.85$&$12.9^{+5.0}_{-3.8}$&$13.0^{+4.9}_{-2.4}$&$8.2^{+3.3}_{-2.0}$&$8.2^{+1.7}_{-2.0}$&$0.27^{+0.34}_{-0.22}$&$0.24^{+0.36}_{-0.21}$&$0.33^{+0.53}_{-0.30}$&$0.36^{+0.50}_{-0.32}$&$0.65^{+0.32}_{-0.78}$&$0.54^{+0.42}_{-1.13}$&$0.18^{+0.73}_{-1.02}$&$0.40^{+0.55}_{-1.17}$\\ 
GW191222$\_$033537&$0.96$&$67.1^{+14.8}_{-9.7}$&$67.2^{+14.7}_{-9.5}$&$52.5^{+10.9}_{-15.4}$&$52.4^{+10.7}_{-15.4}$&$0.35^{+0.52}_{-0.32}$&$0.38^{+0.50}_{-0.34}$&$0.44^{+0.48}_{-0.40}$&$0.41^{+0.50}_{-0.38}$&$-0.18^{+0.99}_{-0.72}$&$-0.16^{+0.97}_{-0.73}$&$-0.07^{+0.93}_{-0.83}$&$-0.09^{+0.95}_{-0.81}$\\ 
GW191230$\_$180458&$0.90$&$82.5^{+19.7}_{-14.7}$&$82.7^{+19.5}_{-13.1}$&$63.1^{+15.5}_{-21.6}$&$62.8^{+13.9}_{-21.4}$&$0.54^{+0.42}_{-0.49}$&$0.51^{+0.44}_{-0.46}$&$0.47^{+0.47}_{-0.42}$&$0.49^{+0.45}_{-0.44}$&$0.01^{+0.84}_{-0.83}$&$-0.10^{+0.92}_{-0.78}$&$-0.28^{+1.06}_{-0.65}$&$-0.16^{+0.99}_{-0.76}$\\ 
GW200112$\_$155838&$0.95$&$44.0^{+8.3}_{-5.4}$&$44.0^{+8.2}_{-5.2}$&$35.2^{+5.5}_{-7.4}$&$35.2^{+5.1}_{-7.4}$&$0.32^{+0.47}_{-0.29}$&$0.34^{+0.46}_{-0.31}$&$0.42^{+0.47}_{-0.37}$&$0.39^{+0.50}_{-0.34}$&$0.09^{+0.78}_{-0.90}$&$0.14^{+0.75}_{-0.91}$&$0.28^{+0.64}_{-1.02}$&$0.25^{+0.67}_{-1.02}$\\ 
GW200128$\_$022011&$0.80$&$64.3^{+16.8}_{-14.7}$&$65.1^{+16.0}_{-9.4}$&$52.5^{+12.5}_{-14.6}$&$51.2^{+10.3}_{-13.3}$&$0.66^{+0.30}_{-0.55}$&$0.58^{+0.38}_{-0.51}$&$0.42^{+0.50}_{-0.38}$&$0.50^{+0.44}_{-0.45}$&$0.45^{+0.49}_{-0.72}$&$0.31^{+0.60}_{-0.95}$&$-0.11^{+0.93}_{-0.76}$&$0.12^{+0.78}_{-0.96}$\\ 
GW200129$\_$065458&$0.91$&$40.2^{+12.3}_{-4.1}$&$40.2^{+12.2}_{-3.3}$&$34.1^{+4.2}_{-10.9}$&$34.1^{+3.3}_{-10.8}$&$0.58^{+0.38}_{-0.50}$&$0.53^{+0.42}_{-0.47}$&$0.45^{+0.46}_{-0.40}$&$0.49^{+0.44}_{-0.42}$&$0.12^{+0.70}_{-0.88}$&$0.13^{+0.74}_{-0.92}$&$0.44^{+0.51}_{-1.27}$&$0.41^{+0.54}_{-1.21}$\\ 
GW200202$\_$154313&$0.95$&$11.0^{+3.8}_{-1.7}$&$11.0^{+3.8}_{-1.5}$&$8.0^{+1.4}_{-1.8}$&$8.0^{+1.2}_{-1.8}$&$0.21^{+0.42}_{-0.19}$&$0.22^{+0.45}_{-0.20}$&$0.35^{+0.52}_{-0.32}$&$0.33^{+0.54}_{-0.30}$&$0.21^{+0.70}_{-0.99}$&$0.17^{+0.73}_{-0.95}$&$0.18^{+0.71}_{-0.92}$&$0.22^{+0.69}_{-0.96}$\\ 
GW200208$\_$130117&$0.95$&$52.9^{+12.2}_{-8.7}$&$52.9^{+12.2}_{-8.3}$&$38.6^{+8.9}_{-11.3}$&$38.5^{+8.6}_{-11.3}$&$0.33^{+0.52}_{-0.30}$&$0.36^{+0.51}_{-0.32}$&$0.46^{+0.47}_{-0.41}$&$0.43^{+0.49}_{-0.39}$&$-0.23^{+1.03}_{-0.69}$&$-0.24^{+1.02}_{-0.68}$&$-0.17^{+1.00}_{-0.74}$&$-0.15^{+1.00}_{-0.76}$\\ 
GW200209$\_$085452&$0.53$&$50.2^{+17.1}_{-15.2}$&$55.7^{+15.0}_{-9.9}$&$48.5^{+17.0}_{-17.3}$&$42.9^{+11.0}_{-13.6}$&$0.67^{+0.29}_{-0.53}$&$0.52^{+0.44}_{-0.46}$&$0.34^{+0.54}_{-0.31}$&$0.49^{+0.45}_{-0.44}$&$-0.52^{+0.70}_{-0.43}$&$-0.24^{+0.97}_{-0.67}$&$0.09^{+0.77}_{-0.90}$&$-0.30^{+1.08}_{-0.63}$\\ 
GW200216$\_$220804&$0.90$&$84.2^{+28.6}_{-25.0}$&$84.4^{+28.4}_{-21.1}$&$51.1^{+25.1}_{-31.3}$&$50.7^{+22.5}_{-30.9}$&$0.53^{+0.42}_{-0.47}$&$0.48^{+0.46}_{-0.43}$&$0.47^{+0.47}_{-0.42}$&$0.51^{+0.44}_{-0.46}$&$0.33^{+0.60}_{-1.05}$&$0.24^{+0.68}_{-1.02}$&$0.06^{+0.85}_{-0.93}$&$0.18^{+0.75}_{-1.03}$\\ 
GW200219$\_$094415&$0.95$&$58.6^{+13.5}_{-9.3}$&$58.6^{+13.5}_{-8.9}$&$44.4^{+9.7}_{-14.2}$&$44.4^{+9.3}_{-14.1}$&$0.49^{+0.45}_{-0.44}$&$0.47^{+0.46}_{-0.43}$&$0.47^{+0.47}_{-0.42}$&$0.48^{+0.46}_{-0.43}$&$-0.28^{+0.96}_{-0.64}$&$-0.23^{+0.96}_{-0.69}$&$-0.09^{+0.93}_{-0.82}$&$-0.17^{+0.99}_{-0.75}$\\ 
GW200224$\_$222234&$0.79$&$51.8^{+9.6}_{-10.0}$&$52.3^{+9.1}_{-5.4}$&$43.8^{+8.0}_{-10.9}$&$42.8^{+5.9}_{-9.9}$&$0.53^{+0.39}_{-0.46}$&$0.46^{+0.45}_{-0.41}$&$0.36^{+0.52}_{-0.33}$&$0.44^{+0.48}_{-0.39}$&$0.46^{+0.47}_{-0.69}$&$0.28^{+0.62}_{-0.96}$&$-0.09^{+0.87}_{-0.79}$&$0.18^{+0.71}_{-1.01}$\\ 
GW200225$\_$060421&$0.88$&$23.4^{+5.7}_{-4.6}$&$23.6^{+5.6}_{-3.2}$&$17.3^{+4.5}_{-4.7}$&$17.2^{+3.0}_{-4.6}$&$0.64^{+0.31}_{-0.52}$&$0.59^{+0.35}_{-0.51}$&$0.36^{+0.53}_{-0.33}$&$0.43^{+0.49}_{-0.39}$&$-0.38^{+0.68}_{-0.53}$&$-0.31^{+0.85}_{-0.59}$&$-0.09^{+0.94}_{-0.82}$&$-0.22^{+1.04}_{-0.71}$\\ 
GW200302$\_$015811&$0.98$&$48.4^{+10.3}_{-9.8}$&$48.4^{+10.2}_{-9.6}$&$25.5^{+12.5}_{-7.6}$&$25.5^{+12.2}_{-7.6}$&$0.36^{+0.51}_{-0.33}$&$0.37^{+0.51}_{-0.34}$&$0.46^{+0.48}_{-0.41}$&$0.45^{+0.49}_{-0.40}$&$-0.03^{+0.87}_{-0.83}$&$-0.02^{+0.86}_{-0.84}$&$0.13^{+0.78}_{-0.98}$&$0.11^{+0.79}_{-0.97}$\\ 
GW200311$\_$115853&$0.97$&$41.9^{+8.2}_{-4.7}$&$41.9^{+8.2}_{-4.6}$&$34.0^{+4.7}_{-7.4}$&$34.0^{+4.6}_{-7.4}$&$0.39^{+0.48}_{-0.35}$&$0.40^{+0.48}_{-0.36}$&$0.42^{+0.50}_{-0.38}$&$0.41^{+0.51}_{-0.37}$&$-0.05^{+0.88}_{-0.80}$&$-0.08^{+0.88}_{-0.79}$&$-0.07^{+0.90}_{-0.81}$&$-0.03^{+0.88}_{-0.83}$\\ 
GW200316$\_$215756&$0.79$&$15.7^{+12.0}_{-5.7}$&$16.0^{+11.8}_{-3.3}$&$9.7^{+5.4}_{-3.6}$&$9.5^{+2.3}_{-3.4}$&$0.37^{+0.39}_{-0.31}$&$0.32^{+0.36}_{-0.29}$&$0.38^{+0.52}_{-0.34}$&$0.44^{+0.47}_{-0.38}$&$0.62^{+0.35}_{-0.56}$&$0.46^{+0.50}_{-1.02}$&$0.10^{+0.78}_{-0.91}$&$0.36^{+0.57}_{-1.05}$
\\[0.1cm]
\hline\hline
&&&&&&&&&\\[-0.3cm]
\multicolumn{1}{c||}{NS binary} & $\zeta$ & \multicolumn{2}{c|}{$m_1\;[M_\odot]$} & \multicolumn{2}{c|}{$m_2\;[M_\odot]$}  & \multicolumn{2}{c|}{$\Lambda_1$} & \multicolumn{2}{c|}{$\Lambda_2$}  \\[0.1cm]
\cline{1-10}
&&&&&&&&&\\[-0.3cm]
GW170817&$0.92$&$1.48^{+0.15}_{-0.12}$&$1.48^{+0.15}_{-0.09}$&$1.28^{+0.11}_{-0.12}$&$1.28^{+0.09}_{-0.12}$&$234^{+615}_{-212}$&$269^{+776}_{-243}$&$517^{+1114}_{-465}$&$446^{+1145}_{-407}$\\ 
GW190425&$0.94$&$1.81^{+0.17}_{-0.10}$&$1.81^{+0.17}_{-0.09}$&$1.61^{+0.10}_{-0.13}$&$1.61^{+0.08}_{-0.13}$&$271^{+1030}_{-248}$&$294^{+1296}_{-269}$&$499^{+1966}_{-451}$&$450^{+1913}_{-409}$

\end{tabular}

\caption{[On two pages.] GW events from BH binaries and NS binaries up to GWTC-3. For each system, we report the fraction $\zeta$ of posterior samples that are left unchanged when comparing the usual mass-sorting strategy and our machine-learning approach. For both labelings, we list medians and 90\% credible intervals of the marginalized posterior distributions. For BH binaries we consider detector-frame masses $m_{1,2}$, spin magnitudes $\chi_{1,2}$, and spin directions $\cos \theta_{1,2}$. For NS binaries we consider detector-frame masses $m_{1,2}$ and tidal deformabilities $\Lambda_{1,2}$.}
\label{table}
\renewcommand{\arraystretch}{1}
\end{table*}
\end{turnpage}

\newpage

\putbib
\end{bibunit}

\end{document}